  \documentclass[iop]{emulateapj}

\newcounter{twocol}
\usepackage{natbib}
\usepackage{colordvi}
\usepackage[usenames]{color}
\usepackage{lineno}

\usepackage{graphicx}
\usepackage{amssymb}

\newcommand{\Fig}[1]{Fig.~\ref{#1}}

\newcommand{\EQA}{\begin{eqnarray}}
\newcommand{\ENA}{\end{eqnarray}}

\newcommand{\joverB}{{|\jvec|/|\Bvec|}}

\newcommand{\jvec}{{\bf j}}
\newcommand{\rvec}{{\bf r}}
\newcommand{\vvec}{{\bf v}}
\newcommand{\Bvec}{{\bf B}}
\newcommand{\evec}{{\bf e}}

\shortauthors{Moreno-Insertis \& Galsgaard}
\shorttitle{A 3D model of coronal jets and eruptions}

\newcommand{\kmsec}{km s$^{-1}$}
\newcommand{\percubiccm}{cm$^{-3}$}
\newcommand{\percc}{cm$^{-3}$}
\newcommand{\persec}{sec$^{-1}$}
\newcommand{\minusone}{$^{-1}$}

\newdimen\sizefig

\begin{document}
\vbox to 7mm{{\bf \vskip 0mm \vskip -3mm \hskip -2mm  To appear in The
    Astrophysical Journal, Vol 770  (20th June 2013)} \vskip 3mm\vfill}

 \title{Plasma jets and eruptions in solar coronal holes: a 3D flux emergence
   experiment}  

\author{F.~Moreno-Insertis\altaffilmark{1,2},
        K.~Galsgaard\altaffilmark{3}}
\altaffiltext{1}{Instituto de Astrofisica de Canarias, Via Lactea, s/n,
  38205  La Laguna (Tenerife), Spain}
\altaffiltext{2}{Dept.~of Astrophysics, Universidad de La
  Laguna, 38200 La Laguna (Tenerife), Spain}
\altaffiltext{3}{Niels Bohr Institute, University of Copenhagen, Juliane Mariesvej 30, 2100 Copenhagen 
}

\begin{abstract}

A three-dimensional numerical experiment of the launching of a hot and fast
coronal jet followed by several violent eruptions is analyzed in detail.
These events are initiated through the emergence of a magnetic flux rope from
the solar interior into a coronal hole. We explore the evolution of the
emerging magnetically-dominated plasma dome surmounted by a current sheet and
the ensuing pattern of reconnection. A hot and fast coronal jet with
inverted-Y shape is produced that shows properties comparable to those
frequently observed with EUV and X-Ray detectors. We analyze its 3D shape,
its inhomogeneous internal structure, and its rise and decay phases, lasting
for some $15$-$20$ min each. Particular attention is devoted to the
field-line connectivities and the reconnection pattern.  We also study the
cool and high-density volume that appears encircling the emerged dome.  The
decay of the jet is followed by a violent phase with a total of five
eruptions. The first of them seems to follow the general pattern of
tether-cutting reconnection in a sheared arcade, although modified by the
field topology created by the preceding reconnection evolution.  The two
following eruptions take place near and above the strong field-concentrations
at the surface.  They show a twisted, $\Omega$-loop like rope expanding in
height, with twist being turned into writhe, thus hinting at a kink
instability (perhaps combined with a torus-instability) as the cause of the
eruption.  The succession of a main jet ejection and a number of violent
eruptions that resemble mini-CME's and their physical properties suggest that
this experiment may provide a model for the {\it blowout} jets recently
proposed in the literature.

\end{abstract}


\section{Introduction}
\label{sec:introduction}

X-ray and EUV plasma jets are a common occurrence in the solar corona,
especially inside coronal holes and along their periphery.  The first
extended study of collimated plasma flows in X rays was achieved through the
Yohkoh satellite in the 1990's \citep{1996PASJ...48..123S,
  1992PASJ...44L.173S}, albeit with limited spatial resolution (e.g.,
2.5\arcsec\ pixel size for Yohkoh-SXT).  Images of the coronal jets in the
EUV were obtained using the EIT instrument aboard the SOHO mission
\citep{1998ApJ...508..899W} and with the TRACE satellite
\citep{1999SoPh..190..167A}.  Substantial observational progress was obtained
from 2006 onward with the soft X-ray telescope and the EUV spectrometer
aboard the Hinode satellite \citep[XRT and EIS, respectively,
  see][]{2007SoPh..243...19C, 2008SoPh..249..263K} and the EUVI and
coronagraph imagers aboard STEREO. The Hinode mission allowed high space and
time resolution observations of the jets in a large range of temperatures
(e.g., for XRT, between $10^6$ K and $3\,10^7$ K, with $1$\arcsec\ and $30$
sec space and time resolution, respectively). They showed that the jets
appear in large numbers in coronal holes, typically with an inverted-Y
shape and with a set of hot magnetic loops at their base
\citep{2007PASJ...59S.771S, 2007Sci...318.1580C}.  The detailed statistics
compiled by \cite{2007PASJ...59S.771S}, covering on the order of $100$ jet
events, indicates that their rate of occurrence is about $60$ per day; their
lifetime in soft X-rays has a broad distribution between about $5$ and $30$
minutes, peaking at $10$ minutes; the jet velocity obtained through tracking
of intensity changes in successive images ranges from $70$ to about $500$
\kmsec, peaking at about $160$ \kmsec. The detection of much higher
velocities (between $600$ and $1000$ \kmsec, \citealt{2007PASJ...59S.771S};
$800$ \kmsec, \citealt{2007Sci...318.1580C}) probably points to phase motions
at the Alfv\'en speed rather than to actual mass transport velocities.
Observed jet heights between $10$ and $120$ Mm and widths between $6$ and
$10$ Mm were also reported by those authors.

Further insights were gained with the STEREO mission.
\cite{2008ApJ...680L..73P} carried out stereoscopic observations of a
polar coronal jet. By identifying the points on the jet front in EUV
images, they found an impulsive acceleration to more than $300$
\kmsec\ which roughly coincides with the appearance of an inverted-Y
jet. The jet developed a helical structure suggestive of untwisting of
magnetic field lines, perhaps following the development of a kink
instability. \cite{2009SoPh..259...87N} studied jets that extended out
to at least $1.7 R_\odot$ and were observable in the EUV from both the
A and B STEREO spacecraft. They found jets with an inverted-Y shape
frequenly associated with lower-lying coronal loops or bright points;
they also find a minor subclass that have the characteristics of
mini-CME events. A little less than half the events reported by those
authors showed a helical structure. The lifetimes of these EUV jets
ranged from about $5$ to $70$ min, with peaks beteween $20$ to $30$
min for the different wavelenghts: in fact, the histograms depend on
the wavelength, with shorter lifetimes for EUV lines formed at higher
temperatures. This is in line with the even shorter lifetimes found
for X-ray jets mentioned above. On the other hand, when using
coronagraph images, the jets have much longer lifetimes, with the
histogram peaking at around $70$ to $80$ min.  More recently,
\citet{2010ApJ...720..757M} have discussed a subclass of polar coronal hole
X-ray jets (about one third of their sample), which they call {\it
blowout jets}, which are counterparts of erupting-loop H$\alpha$
macrospicules. The members of the subclass are
  characterized by showing both a standard (or quiescent) jet and a
  more violent eruption at their base, possibly indicating that a
flux rope is ejected, as in a mini-CME event
\citep{2010A&A...517L...7I}. These observations are particularly
interesting since, as shown in the current paper, the magnetic
topology resulting from the emergence of magnetic flux into a coronal
hole, subsequent reconnection and jet ejection is complicated enough
for the system to strive to reduce its free energy via sudden eruptive
events. Further instances of this sort of
  behavior have been discussed by \citet{2011ApJ...735L..18L} and
  \citet{2011A&A...526A..19M}.

Concerning plasma properties in the jets, a Yohkoh-based determination was
provided by \cite{2000ApJ...542.1100S}. They found temperatures between $3$
and $8$ million K and densities between $7\,10^8$ and $4\,10^9$
\percubiccm. The recent work of \cite{2012A&A...545A..67M} finds lower
temperatures, $2.5$ million K, and densities around $10^{8.8}$
\percubiccm\ in the jets, and similar temperatures (but higher densities,
$10^{9.5}$ \percubiccm) at the base of the jets, although in a single case
the temperature there reached at least $1.2\,10^7$ K. Further relevant
references for the plasma properties of the jets are
\cite{2007PASJ...59S.751C, 2009A&A...503..991M, 2010A&A...510A.111D,
  2011A&A...535A..95D, 2010ApJ...710.1806D, 2010A&A...516A..50S,
  2011A&A...526A..19M}. All in all, the observations support magnetic
reconnection as the mechanism through which free magnetic energy is released
and turned, via Joule heating, into thermal energy in the current sheet
region, with temperatures reaching near $10^7$ K, as well as into kinetic
energy via the reconnection outflows, shock acceleration and subatomic
particle acceleration.

The theoretical explanation for the jets has proceeded on the basis of
detailed 2D and 3D numerical experiments using as starting point simple
physical ideas about the triggering mechanism. The most explored theoretical
idea is that the jets are produced when new magnetic flux emerges from the
solar interior and collides with magnetic domains with simple (e.g., uniform
field) configurations in the corona. This idea goes back to
\cite{1977ApJ...216..123H}, who proposed it as a mechanism to unleash flare
eruptions. A related 2D numerical experiment was carried out by
\citet{1984SoPh...94..315F}.
The first two-dimensional numerical MHD models of inverted-Y jet ejection
following flux emergence from the solar interior were calculated by
\cite{1995Natur.375...42Y, 1996PASJ...48..353Y}.  Their models, which are
strictly 2D (i.e., without velocity or magnetic field
components in the direction of the ignorable coordinate), confirm that the
interaction of a magnetic loop system emerging from the solar interior with
an open, slanted field in the corona leads to the formation of a current
sheet and ensuing reconnection at the interface between the two systems. The
reconnection outflows, upon impinging onto the nearby open field system, lead
to a fast-mode shock which diverts part of the outgoing plasma upward along
the open field lines, thus creating a jet at one end of the current sheet
(see Fig.~9 in the paper by \citealt{1996PASJ...48..353Y}).  The authors
obtained a jet with velocity $0.6$ times the background coronal Alfv\'en
speed and temperature $1.6$ times the background coronal temperature.  At the
other end of the current sheet, a set of closed coronal loops is created by
the reconnection, which, therefore, also contain hot plasma.  This experiment
was a clear landmark. Yet, to alleviate the timestep restrictions of a
realistic corona, the authors adopted unrealistic values for the density
($\approx 10^{12}$ \percubiccm), temperature ($\approx 2.5\,10^5$ K) and
Alfv\'en velocity ($\approx 84 $ km s$^{-1}$) of the background coronal
model, obtaining jet temperatures and velocities far from the observed
values.  In recent years, \cite{nishizuka_etal_2008} have repeated the
\citeauthor{1996PASJ...48..353Y} experiment, but now with $T = 10^6$ K and
$n=10^{10}$ \percubiccm\ in the corona.
They obtain jet velocities and temperatures closer to the observations. The authors focus on
the possible detection of Alfv\'en waves in the jet and on the
quasi-simultaneous emission of neighboring hot and cold ($10^4$ K) plasma
jets (see Sec.~\ref{sec:discussion}). They also discuss similarities with
X-Ray, EUV and CaII observations.

Various other aspects of coronal jet emission following flux emergence have
been explored the past years with 2D experiments.  The reconnection in the 2D
models may have an oscillatory character, as shown by
\cite{2009A&A...494..329M} through a 2.5D experiment
(i.e., one that allows for non-zero velocity and
  magnetic field components along the ignorable coordinate): the
emerged-field and hot-loop domains of the reconnection site change role as
inflow and outflow regions for the reconnection in a damped oscillatory
fashion following the corresponding increase and decrease of the gas pressure
in them.  \cite{2012ApJ...749...30M} further investigated the
dependence of the oscillations on model parameters.

The first numerical experiment in three dimensions of jet launching
associated with magnetic flux emergence from the interior was reported in a
Letter by \cite{2008ApJ...673L.211M}. Like in the earlier 2D experiments, their
domain stretches from the top of the convection zone to the corona, but now
the corona is endowed with realistic values of temperature ($1.1\,10^6$ K),
density ($\approx 2\,10^8$ \percubiccm) and magnetic field ($10$ G) adequate
for a coronal hole domain. The initial coronal field is uniform and inclined
$25$ deg from the vertical direction. The field emerging from below the
photosphere is a twisted magnetic tube of the $\Omega$-loop kind, as in the
experiments by \citet{2004A&A...426.1047A} or \citet{galsgaardetal07}.  The
experiment showed that both the emerging domain and the hot reconnected
coronal-loop region are dome-shaped, with the current sheet
covering the top of the former. The jet itself had the
canonical inverted-Y profile when seen in 
perspective from most orientations around the dome. High velocities (up to
$200$ km s$^{-1}$) and temperatures (slightly above $10^7$ K) were measured
in the jet, but it was not reported how volume-filling and permanent the
high-T and high-velocity regions were in the jet evolution. A high-density
region ($10^{10}$ \percubiccm) was found outside of the emerging dome roughly
underneath the jet: it contains plasma formerly located in the emerged dome
that has been dumped onto the open-field domain as part of the reconnection
process. The paper by \cite{2008ApJ...673L.211M} also contained a comparison
with Hinode/XRT and EIS observations of a hot coronal jet.

The aim of the current paper is to describe in detail a three-dimensional
experiment of jet emission resulting from flux emergence from the solar
interior that extends the work of \citet{2008ApJ...673L.211M}.  Major
differences to that paper are that our domain here is much larger, including
a coronal volume of  $68$ Mm height, and that we pursue the experiment for an
extended time, namely, about $80$ minutes after initiation of the buoyancy
instability leading to emergence into the atmosphere, which is well beyond
the decay phase of the hot jet. We provide descriptions of the
initial phases of the evolution, the current sheet, the jet, and the system
of reconnected hot loops. The dense domain generated as a sub-product of the
reconnection is also studied including Lagrange-tracing of the
plasma elements.  Further, a series of violent eruptions starting at
different positions of the emerged flux rope could be detected: we count five
eruptions in total, all taking place after the jet was already in the
advanced decay phase. This links our experiment with a number of flux
emergence simulations of the past ten years which focused on flux-rope
eruptions \citep{2004ApJ...610..588M, 2008A&A...492L..35A,
  2008ApJ...674L.113A, 2012A&A...537A..62A, 2009A&A...508..445M}, and also
with the experiments of \citet{2003A&A...406.1043T, 2005ApJ...630L..97T,
  2009ApJ...691...61P, 2010ApJ...714.1762P} and \citet{2010ApJ...708..314A},
who obtained eruptions following kink- or torus-instability patterns not
associated with magnetic flux emergence from the interior.  The first of the
flux emergence references given above \citep{2004ApJ...610..588M} describes
how the emerged domain is subjected to shear across the main polarity
inversion line (PIL) turning into a sheared arcade. 
A vertical current sheet is formed and reconnection ensues, following a
pattern not completely unlike the classical mechanism of
\cite{1964NASSP..50..451C}, \cite{1966Natur.211..695S},
\cite{1974SoPh...34..323H}, and \cite{1976SoPh...50...85K} for the launching
of a flare: a flux rope is violently ejected upward containing the
reconnected top part of the original arcade; simultaneously, a sort of {\it
  post-flare} magnetic arcade is created at the bottom end of the sheet.  In
our experiment, we obtain a first eruption centered between the
opposite-polarity roots of the emerged domain that follows this general
pattern even if importantly modified through the different coronal field
configuration prior to the flux emergence.  The subsequent eruptions,
however, take place on the sides of the emerged active regions and 
may contain a combination of the kink and torus instabilities in them.

The layout of the present paper is as follows. In Section~\ref{sec:model} we
provide details of the numerical model. Section~\ref{sec:structure} discusses
the initial evolution and the structure after emergence. In
section~\ref{sec:jet_reconnection} the main phase of the jet activity is
presented, including details of the current sheet, the reconnection and the
jet itself, as well as of the closed loop system resulting from the
reconnection. Section~\ref{sec:cold_region} describes the formation of a cold
density region encircling the emerged domain.
Section~\ref{sec:advanced_phases} describes the series of repeated eruptions
that take place after the decay of the jet, including details of the magnetic
connectivity changes that result from them.
Section~\ref{sec:time_evolution_global} looks into the evolution of global
quantities, like the axial magnetic flux or the interior-to-corona
magnetic connectivity.
 Finally, section~\ref{sec:discussion}
contains a summary and discussion.

\section{Model and equations}
\label{sec:model}

\begin{figure}
\sizefig=16.cm \ifnum \value{twocol} > 0 \sizefig=9.5cm \fi
\centerline{\hfill
 \includegraphics[width=0.5\textwidth]{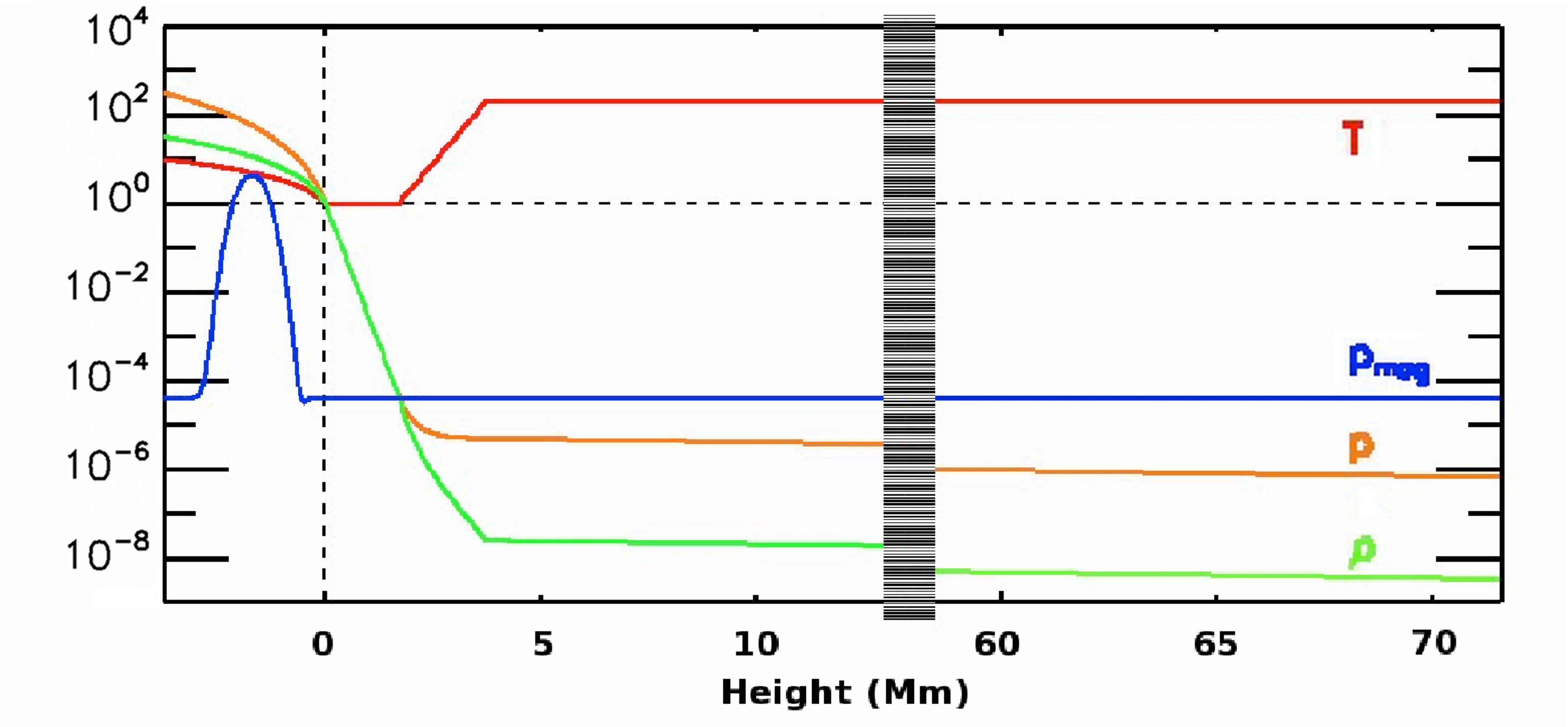}\hfill}
\caption[]{Stratification in the initial model. Shown are the temperature
  (red), gas pressure (orange) and density (green) as a function of height
  height at $t=0$ far from the magnetic tube, normalized to their
  photospheric value.  The blue line is the magnetic pressure distribution
  along a vertical line passing through the center of the initial tube in the
  same units as the gas pressure.  For clarity, the central sections of the
  corona have been omitted.
\label{fig:stratif}}
\end{figure}

We model the emergence of a twisted magnetic
flux rope from underneath the solar surface into an open-field domain in the
corona. Three basic structural elements at the initial time are, therefore:
(a) the background stratification; (b) the ambient coronal magnetic field;
and (c) the subphotospheric magnetic tube. The initial background model is
similar to that used in our previous papers on 3D flux emergence
models \citep[see][]{2004A&A...426.1047A}, but with a very extended
corona. In Fig.~\ref{fig:stratif} we present the dependence of temperature
($T_{st}$, red), density ($\rho_{st}$, green) and pressure ($p_{st}$,
orange) with height ($z$) at time $t=0$, all normalized to their
photospheric ($z=0$) values. The domain consists of four regions in
hydrostatic equilibrium representing, from bottom to top, the uppermost
$3.74$ Mm of the convection zone (assumed adiabatically stratified), the low
atmosphere (a domain between $z=0$ and $z=1.70$ Mm initially with uniform
temperature, $T_{phot}$),
a transition region between $z=1.70$ and $z=3.74$ Mm, and above it the
corona, that extends for $67.9$ Mm and 
has $T= 200\;T_{phot}$.  The units of temperature, density and pressure for
the calculation are taken to be the photospheric values, namely:
$T_{phot}=5.6\,10^3$~K, $\rho_{phot} = 3\,10^{-7}$~g~\percc\ and
$p_{phot}=1.4\,10^5$~erg~\percc. This leads to the following units of
velocity: $c_{phot} =
\sqrt{p_{phot}/\rho_{phot}} = 6.8$ km \persec; length: $H_p =
p_{phot}/(\rho_{phot}\,g_\odot) = 170$ km; and time: $\tau= 25$ sec.  The
coronal values for temperature and particle density that follow from those
choices are: $T_{cor}=1.1\,10^6$ K and $n_{part}$ between $5.9\,10^8$ \percc\
and $4.3\,10^9$ \percc, respectively, which are reasonable values for a
coronal hole atmosphere \citep{2006A&A...455..697W,2002SoPh..211...31H}. Like
in the previous papers, an elementary ideal gas with uniform
specific heats is used here for simplicity. 

To mimic a coronal hole medium, the corona must be endowed with an open
magnetic field. To that end, we add in the whole box a uniform ambient
magnetic field of $10$ G intensity contained in the $x-z$ plane
and inclined an angle $\alpha=25$ deg to the vertical, 
$\Bvec =
B_c\,( -\sin\alpha\; \evec_x + 
\cos \alpha\; \evec_z)$, with $\evec_i$ indicating the basic unit vectors and
$B_c$ the field strength.
This is of course a potential field, so the original hydrostatic
equilibrium is preserved. The associated plasma $\beta$ is small in the
corona (between $0.12$ and $0.02$); on the other hand, this field is
dynamically irrelevant in the photosphere and below, where $\beta > 3\,10^4$.

The initial setup is completed by embedding a horizontal, twisted
magnetic flux rope oriented along the $y$-axis below the photosphere at
$z=-1.7$ Mm.  Like in the earlier experiments, the prescription for this
field follows \cite{2001ApJ...554L.111F}: using cylindrical coordinates
$(r,\theta,y)$ around the tube's axis, the field components are given by:
\begin{equation}
B_r=0 \,; \quad B_\theta = q\,r\,B_y\,; \quad B_y = B_0\,\exp(-r^2/R^2) \,;
\label{eq:field_components}
\end{equation}
with $q$ a constant twist parameter.  This yields a single value of the
pitch, $2\pi/q$, for all field lines. Here we are using $B_0 = 3.8$ kG, $R =
0.4$ Mm, and $q=2.4$ Mm\minusone, leading to a total axial magnetic flux of
$2.2\,10^{19}$ Mx, which is in the range of the ephemeral active regions. The
choice of sign for $q$ is such that it leads to opposition between the
direction of the $B_\theta$ component of the tube and the ambient coronal
field described above.
The magnetic pressure distribution for the total (tube +
ambient) field along a vertical line that intersects the tube axis can be
seen as a blue solid line in Fig.~\ref{fig:stratif}.  The gas pressure is
then modified from $p = p_{st}$ to $p = p_{st} + p'$ in such a way that the
sum of Lorentz and gas pressure forces, ${\bf F}_{Lor} - \nabla\,p$, has the
same value as before introducing the magnetic tube; hence, the hydrostatic
equilibrium is preserved.
This choice is always possible since the
Lorentz force for the given field is curl-free. Finally, the horizontal
magnetic tube is made buoyant by increasing the entropy of its plasma
elements for a limited distance in $y$ around $y=0$ but without modifying the
pressure. The resulting density perturbation, $\rho' = \rho - \rho_{st}$, is
of the form $\rho'/\rho_{st} = (p'/p_{st})\; \exp(-y^2/\Lambda^2) \le 0$. We
choose $\Lambda = 3.4$ Mm.

The experiment is solved in a Cartesian domain of size $48.3$ Mm x $33.9$ Mm
x $75.4$ Mm in $(x,y,z)$, respectively.  This domain is substantially taller
than in all our previous 3D flux emergence experiments, and includes, in
particular, $71.6$ Mm above the photosphere. This is done so that the we can
capture a substantial distance along the jet and to separate completely the
top boundary domain from any region of interest in the box.  Concerning the
boundary conditions, the domain is
taken to be periodic in the horizontal directions. 
The boundaries in the vertical direction are impenetrable and with zero
horizontal velocity. This basic setup would reflect any propagating
perturbation hitting the top and bottom boundaries, which would be
undesirable, especially since a jet is meant to propagate upward in the box. 
Hence, a non-reflecting damping
layer is implemented near those boundaries: 
we use an algorithm of the Newton-cooling type that brings the
density, internal energy and momenta back toward their $t=0$ values
on a timescale that increases with distance from the
boundary. 

The evolution of the experiment is followed in time by solving the non-ideal
MHD equations (as given in the paper by \citealt{2004A&A...426.1047A}) using
the Copenhagen Stagger Code which contains
  hyperdiffusive ohmic and viscous terms
self-consistently included in the induction, momentum and energy equations
 \citep{Nordlund_Galsgaard1997}.
The numerical box used has $448$ x $320$ x $512$ points in the $(x,y,z)$
directions, respectively. The grid is uniform in the horizontal directions
with about $100$ km resolution. In the vertical direction, we use a stretched
grid to better resolve the low atmosphere and the reconnection site that
produces the jet. We have a vertical resolution of $30$ km at the
photospheric level and of $100$ km at a height of $10$ Mm above the
photosphere.

\section{Initial phases and structure after emergence}
\label{sec:structure}

\begin{figure}
\sizefig=7.7cm \ifnum \value{twocol} > 0 \sizefig=8.2cm \fi
\vbox{
 \centerline{\includegraphics[width=\sizefig]{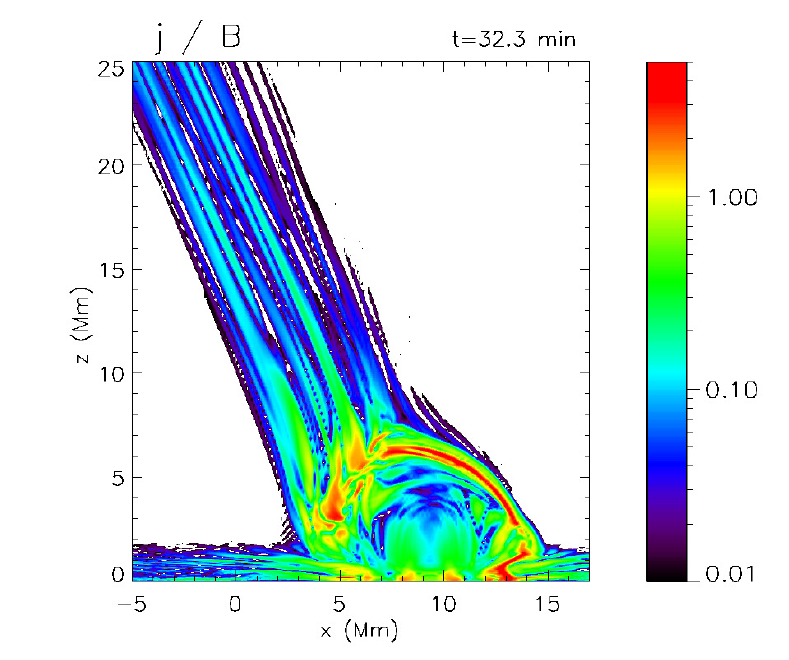}}
\vspace{2mm}
 \centerline{\includegraphics[width=\sizefig]{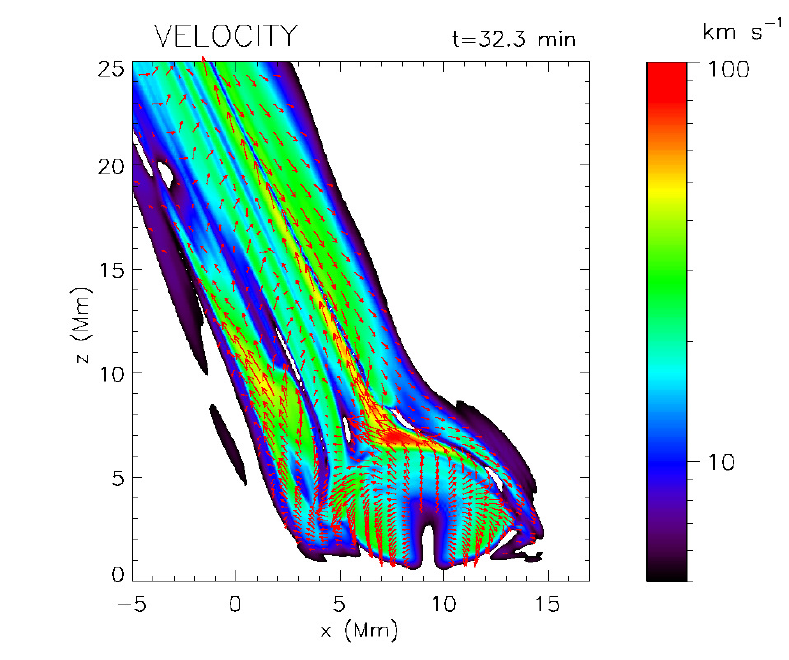}}
\vspace{2mm}
 \centerline{\includegraphics[width=\sizefig]{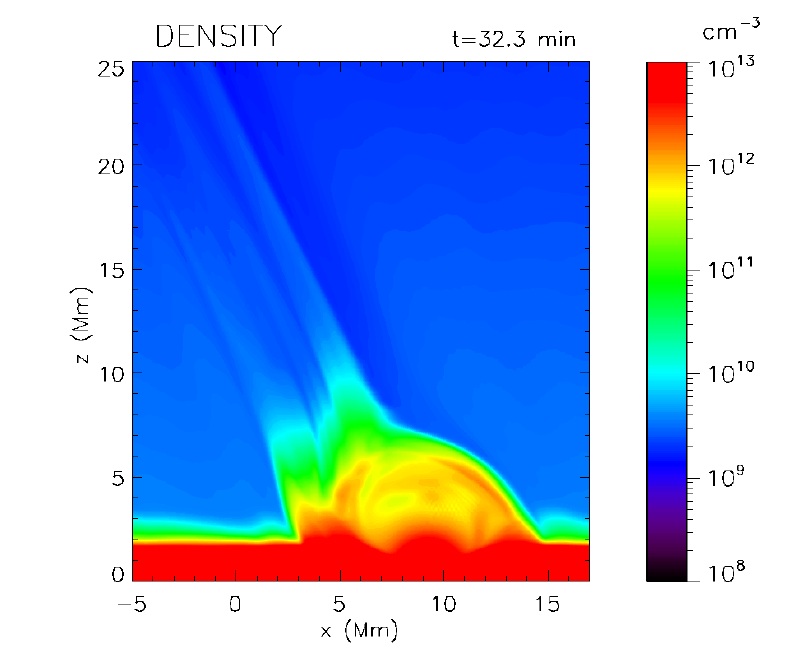}}
}
\caption[]{ 2D vertical cut across the box at $y=0$, i.e., perpendicular to
  the initial axis of the tube. Shown are color maps of $|\jvec|/|\Bvec|$
  (top), $|\vvec|$ (center) and $\rho$ (bottom) at an early time
  of the jet evolution, $t=32.3$ min. Variable values below the minimum of
  the colorbar are drawn in white. The current sheet at the periphery of
  the emerged dome is visible as a thin red stripe (top panel). The maximum
  velocity ($82$ km \persec) is reached in the reconnection region. 
\label{fig:2Dcut_early}}
\end{figure}

The initial stages of the evolution share many common features with previous
papers based on similar initial conditions \citep[like,
  e.g.,][]{2004A&A...426.1047A, archontis_etal_05, galsgaardetal07,
  2008ApJ...673L.211M, 2009A&A...494..329M}.  The central part of the tube,
which, following the prescriptions explained in Sec.~\ref{sec:model}, has a
small density deficit compared with its surroundings, $|\rho'/ \rho_{st} | =
$O$(0.04)$, rises toward the photosphere leading to an $\Omega$-loop shape
for the tube.  The rise to the surface takes some $10$ minutes.  The
temporary slowing down in those levels and the development of a secondary
buoyancy instability there take a further $10$ min, approximately, and are as
described or analyzed by different authors (e.g., for the secondary buoyancy
instability, \citealt{magara01} in 2D and \citealt{2004A&A...426.1047A},
\citealt{Moreno-Insertis:japan06} and \citealt{murray_etal_2006} in 3D).
This buoyancy instability leads to the runaway rise of magnetized plasma. By
time $t=25$ min, for instance, one can see a roundish hill of strongly
magnetized material pushing into the corona with its summit located about $3$
Mm above the photosphere (as in
  Fig.~\ref{fig:2Dcut_early}, bottom panel, but less expanded than in that
  more advanced time) and showing an abrupt drop of magnetic pressure
between the emerged material and the ambient corona. By time $t=27$ min,
i.e., still quite early in the evolution, the drop has turned into a smooth
transition and the emerged system, while still magnetically dominated [$\beta
  \sim \hbox{O}(10^{-2})$], enters a slower rising phase, characterized by a
quasi-equilibrium between the outward-pointing magnetic pressure gradient
force and the inward-pointing magnetic tension force.  This approximately
force-free phase is possible since the Alfv{\'e}n crossing time is short,
below $1$ min, well below the evolutionary time scale dictated by the
instability development. During this phase the magnetic field strength in the
emerged dome decreases with height following quite accurately an exponential
law, with scaleheight $H_B \approx 1.5$ Mm at $t=30$ min and $H_B \approx
2.3$ Mm at $t=40$ min. This is reminiscent of the self-similar expansion
evolution described by \cite{1989Apj...338..471S} for the 2D development of a
buoyancy instability in a magnetic layer.

In these early stages the emerged plasma is very dense compared to the
initial unperturbed values at the same height: $\rho$ values
corresponding to heights between $1$ and $2$ Mm in the initial
atmosphere appear now on a much larger range of heights (e.g., up to $6$
Mm by time $t=30$ min). Yet, gravity is small compared with the components
of the Lorentz force in this region:
the $H_B$ values quoted above yield ratios $v_A^2 / H_B >> g_\odot^{}$, so
that the plasma can only fall along the field lines. 
The velocity pattern in the emerged plasma in this phase is then
comparatively simple: still propelled by the buoyancy instability, the
rising mountain continues its advance into the atmosphere with an almost
force-free magnetic field configuration. The plasma in it slides down
along the field lines thereby reaching velocities on the order
of the free-fall value (e.g., about $40$ km s$^{-1}$ for an element
falling under gravity for a distance of about $5$ Mm). When seen in a
two-dimensional cut, the velocity field pattern has a characteristic
fountain-like shape (still visible at a later time
through the arrows in the dome-shaped domain  
of Fig.~\ref{fig:2Dcut_early}, central panel).

Soon after entering the atmosphere, a prominent current sheet is formed at
the periphery of the emerging plasma mountain (as visible in the form of a
red stripe around the dome in Fig.~\ref{fig:2Dcut_early}, top panel, for
$t=32$ min). This current sheet directly results from
  the conflict of orientations of the azimuthal magnetic field component of
  the rising tube with the ambient coronal field. In 3D one can see that the
  sheet covers, as a blanket, all that side of the rising mountain. 
Reconnection sets in across the current sheet from the early stages of the
emergence and is already well under way at the time shown in
Fig.~\ref{fig:2Dcut_early} ($t=32$ min). 
The central panel of the figure shows a color map of $|\vvec|$, with,
superimposed as red arrows, the projection of the velocity vector on the
$x-z$ plane.  The plasma is being ejected from the current sheet toward the
ambient corona on the left of the figure. The inflow into the sheet both from
the corona and from below, i.e., from the emerged dome, is also
apparent.  Feeding the current sheet from the corona, in particular, creates
a downflow from high up along the coronal field lines adjacent to the
reconnection site. At this early stage, however, the reconnection-related
velocities are not large: several tens of km sec$^{-1}$ in the coronal downflow,
and up to $100$ km sec$^{-1}$ in the reconnection outflow. It is also visible in
the figure that there is an ejection of plasma (hence, of reconnected field
lines) further out to the left (e.g., for $x < 3$ Mm). Those field lines
carry the dense material from the 
emerged mountain with themselves, producing a cool, dense region next to the
emerged dome: the green, spike-like area to the left of the reconnection site
in the central panel [e.g. $(x,z) \sim (2,8)$ Mm] might be taken as a sort of
cool jeet with velocities of 
several 10s of km sec$^{-1}$. Yet, in fact, when seen in 3D, one realizes that
the cool region encircles the emerged dome and has no obvious jet
shape. Details are given in Sec.~\ref{sec:cold_region}. The bottom panel of
Fig.~\ref{fig:2Dcut_early} will be discussed in the next section.

\section{The main phase of jet activity} 
\label{sec:jet_reconnection}

\begin{figure}
\sizefig=7.5cm \ifnum \value{twocol} > 0 \sizefig=8cm \fi
\centerline{\hfill
\vtop{
\centerline{\includegraphics[width=\sizefig]{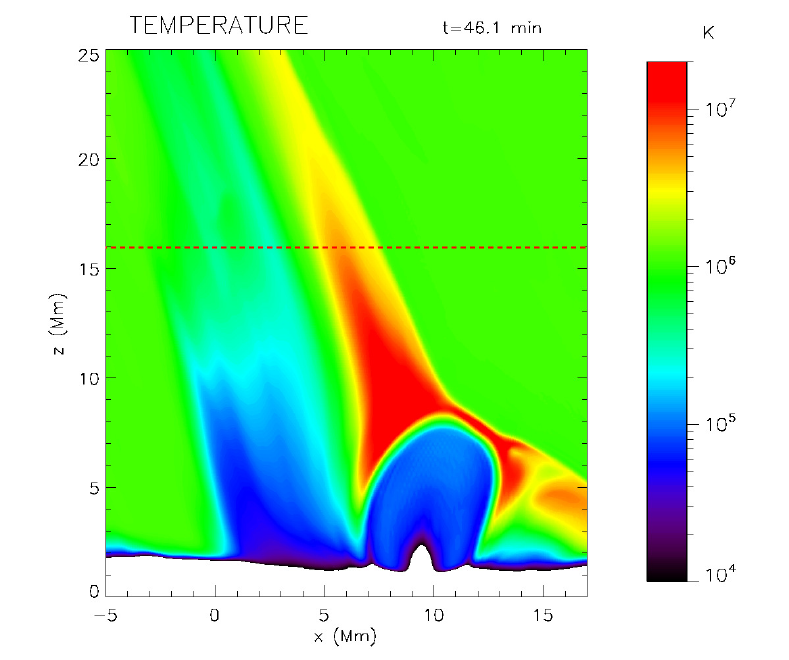}}
\vspace{2mm}
 \centerline{\includegraphics[width=\sizefig]{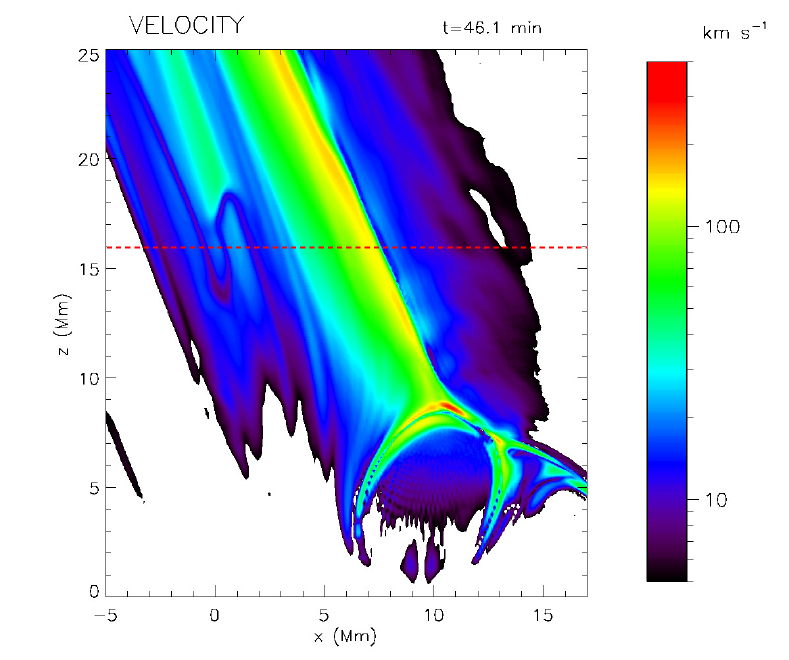}}
\vspace{3mm}
 \centerline{\includegraphics[width=\sizefig]{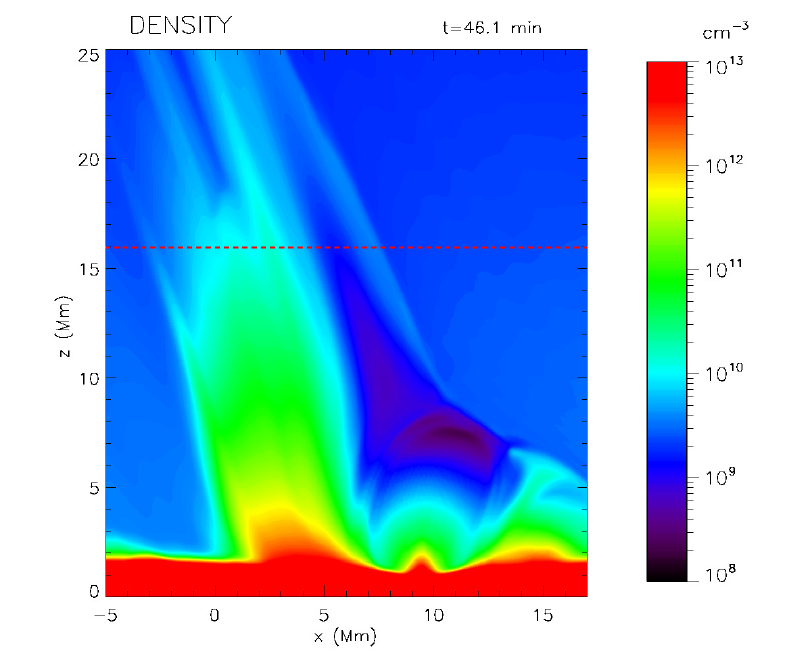}}
}}
\caption[]{2D maps for the same vertical plane as for
  Fig.~\ref{fig:2Dcut_early} toward the peak of the jet activity ($t=46$
  min). The panels here correspond to temperature (top), $|\vvec|$ (center),
  and density $\rho$ (bottom). Variable values below the minimum of the colorbar are
  drawn in white. The maximum values of temperature ($2.2\,10^7$ K) and
  velocity ($290$ km \persec) in the domain shown are reached in the
  reconnection region. The red dashed line indicates the height used for
  Fig.~\ref{fig:jet_horizontal_cuts} 
\label{fig:2Dcut_full}}
\end{figure}

\begin{figure}
\sizefig=8.cm \ifnum \value{twocol} > 0 \sizefig=7.5cm \fi
\centerline{\hfill
\includegraphics[width=\sizefig]{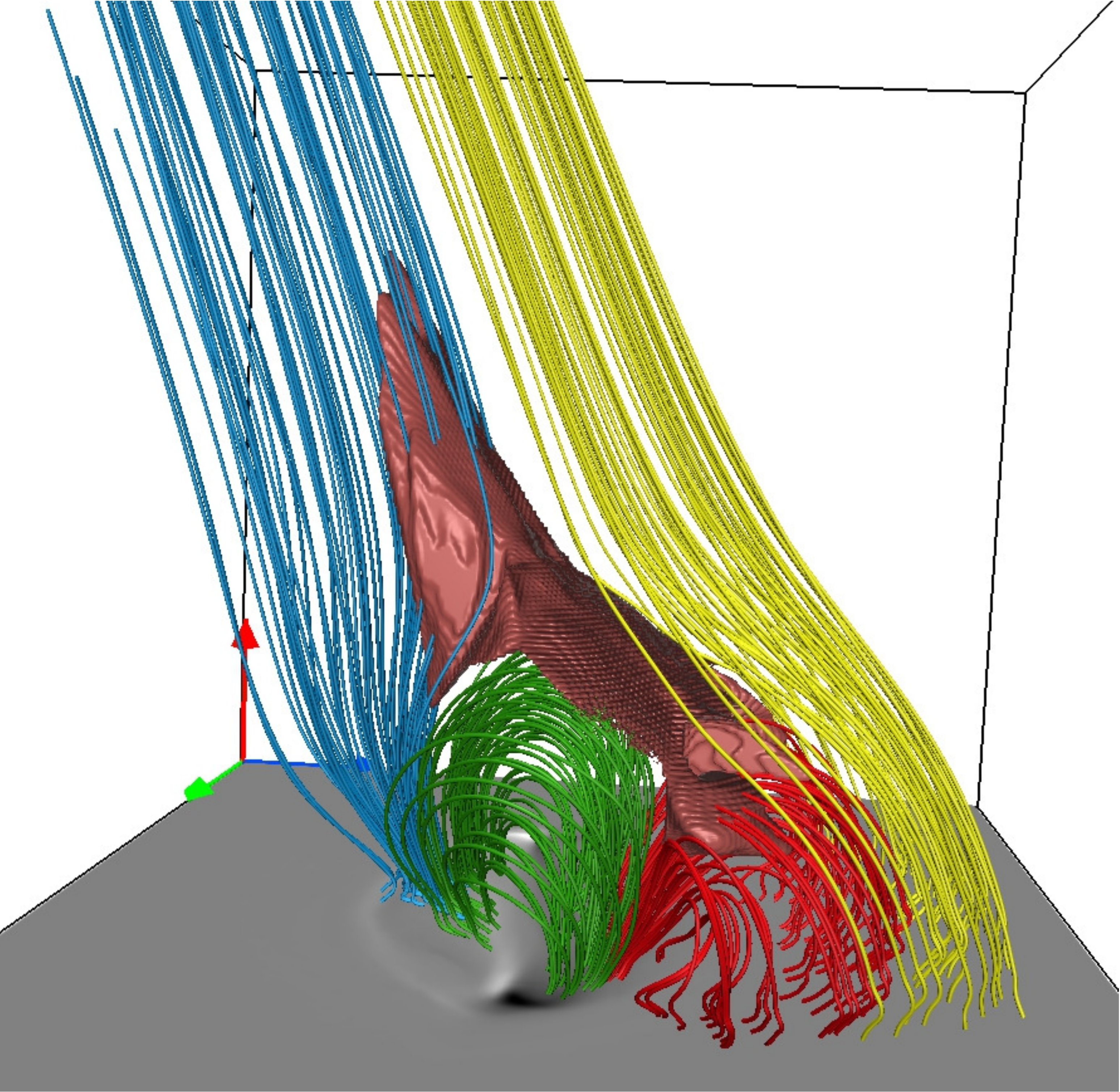}\hfill}
%
%
\caption[]{3D view of the jet (visualized through a temperature isosurface)
  and field lines of the different flux systems}
\label{fig:3Dview}
\end{figure}

Following the standard picture, the reconnection taking place across the thin
current sheet lets magnetic field and plasma from the emerged system join the
field and plasma coming down from the corona and ejects them toward either
end of the sheet. The situation is comparatively easy to describe by using a
vertical cut of the system in the x-z direction, as in
Figs.~\ref{fig:2Dcut_early} and \ref{fig:2Dcut_full}. The reconnected field
lines ejected toward the left-hand side of the figure become open field
lines rooted in the photosphere. The reconnection outflow in that direction
leads to the creation of an upward moving jet along the field lines, which is
studied in detail in subsection \ref{sec:the_jet} below. It also leads to
flows pointing toward the photosphere along the left flank of the emerged
dome; their discussion is deferred to Sec.~\ref{sec:cold_region}.  The
reconnection outflows that point toward the right-hand side of the current
sheet in the figure lead to the formation of a system of coronal
loops (i.e., with both footpoints rooted in the photosphere) which are 
studied in Sec.~\ref{sec:hot_loops}.
Finally, the {\it jet engine}, the reconnection in the current sheet, is
studied in Sec.~\ref{sec:current_sheet_reconnection}.

\subsection{Time evolution of the jet}\label{sec:the_jet}

In the early phase, the material in the emerged system is very dense
(Fig.~\ref{fig:2Dcut_early}, bottom panel), with $\rho$ being above the
density of the ambient coronal plasma at the same height by one or two orders
of magnitude. Consequently, the reconnection outflows and the ensuing jet
strands do not have high temperature.  However, as a consequence of the
continued drainage of plasma along the field lines explained in the previous
section, the material in the emerged region becomes increasingly rarefied.
Hence, at the peak of the reconnection process (Fig.~\ref{fig:2Dcut_full}),
we find that an important part of the plasma flowing into the current sheet
from below has a low density (Fig.~\ref{fig:2Dcut_full}, bottom panel), even
down to an order of magnitude below the ambient coronal values. The ohmic
heating taking place in the current sheet, therefore, can raise their
temperature to high levels (Fig.~\ref{fig:2Dcut_full}, top panel), up to
O($10^7$ K) in the neighborhood of the sheet. The result, seen in the 2D
projection of Fig.~\ref{fig:2Dcut_full}, is a triangular region (actually, of
inverted-Y shape) with low $\rho$ and high $T$ values in the lowermost $\sim
10$ Mm of the jet. The time evolution of the temperature in that vertical
plane can be studied through an accompanying movie.  A 3D view is provided in
Fig.~\ref{fig:3Dview}: the figure includes a magnetogram at the photosphere
showing the strong negative surface polarity at the front; there is also a
temperature isosurface (brown) at $T=7\,10^6$ K, that contains in its
interior the reconnection site and delineates the beginning of the jet and
the top of the hot loops formed after reconnection. The
  four field line systems visible in the figure are described in
  Sec.~\ref{sec:current_sheet_reconnection}.

\begin{figure}
\sizefig=6.cm \ifnum \value{twocol} > 0 \sizefig=5cm \fi
\centerline{
 \includegraphics[width=\sizefig]{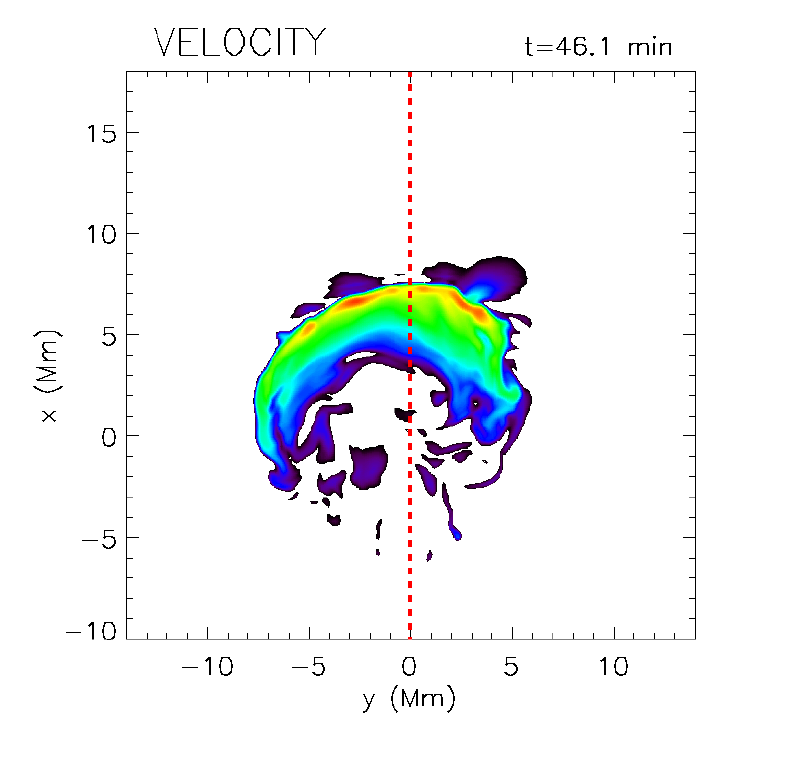}\hskip -2mm
 \includegraphics[width=\sizefig]{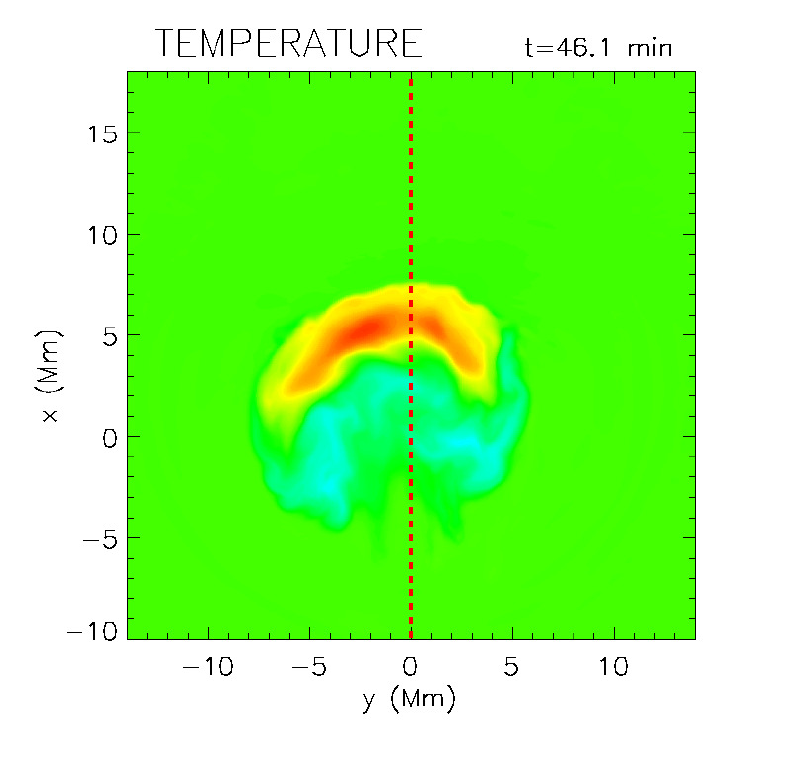}}
\vspace{-2mm}
\centerline{
 \includegraphics[width=\sizefig]{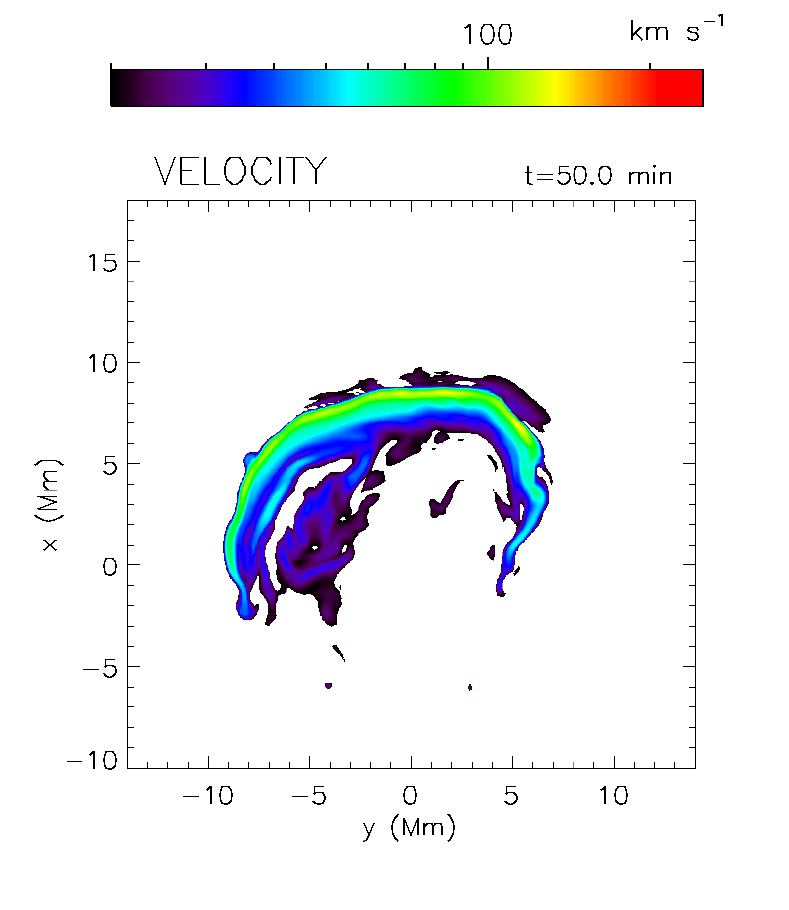}\hskip -2mm
 \includegraphics[width=\sizefig]{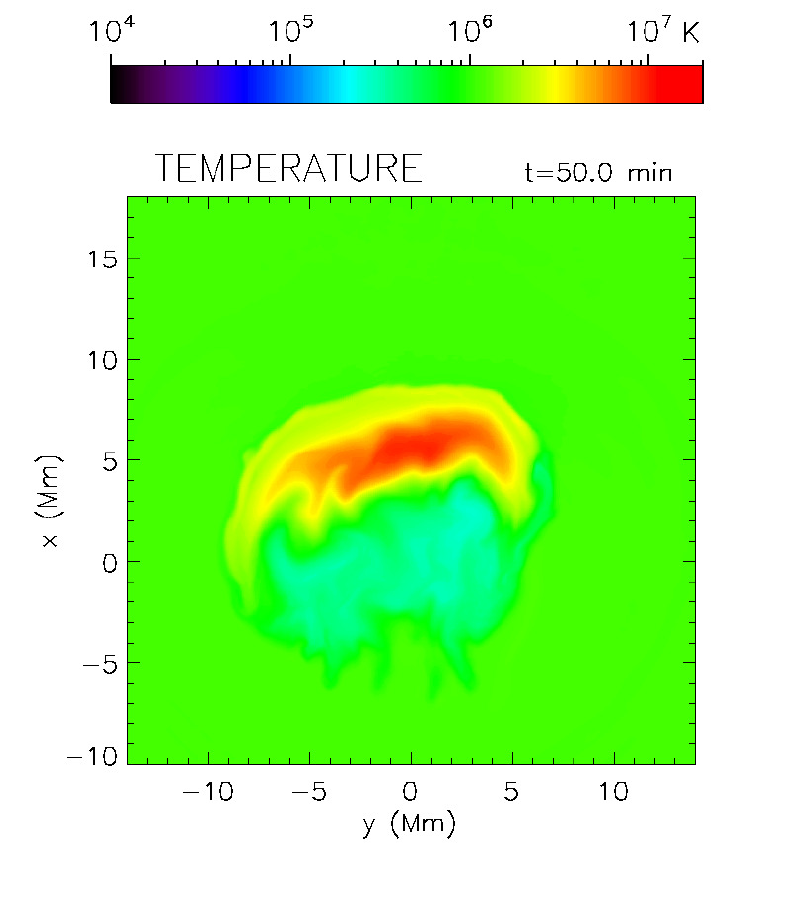}}
\caption[]{Color maps of velocity (left) and temperature (right) on a
  horizontal cut taken at a height of $16$ Mm above the photosphere cutting
  across the jet. Top row: $t=46.1$ min (simultaneous with
  Fig.~\ref{fig:2Dcut_full}). Bottom row: $t=50.0$ min. A number of 
  high-velocity strands are visible in the top-left panel. The jet is seen as
  a bent surface containing different layers: an external high-velocity
  sheet, and a more internal hot layer. 
  The red dashed line in the top panels indicates the location of the cut in
  Fig.~\ref{fig:2Dcut_full} 
\label{fig:jet_horizontal_cuts}}
\end{figure}

\begin{figure}
\sizefig=14.cm \ifnum \value{twocol} > 0 \sizefig=8cm \fi
\centerline{\hfill
\includegraphics[width=\sizefig]{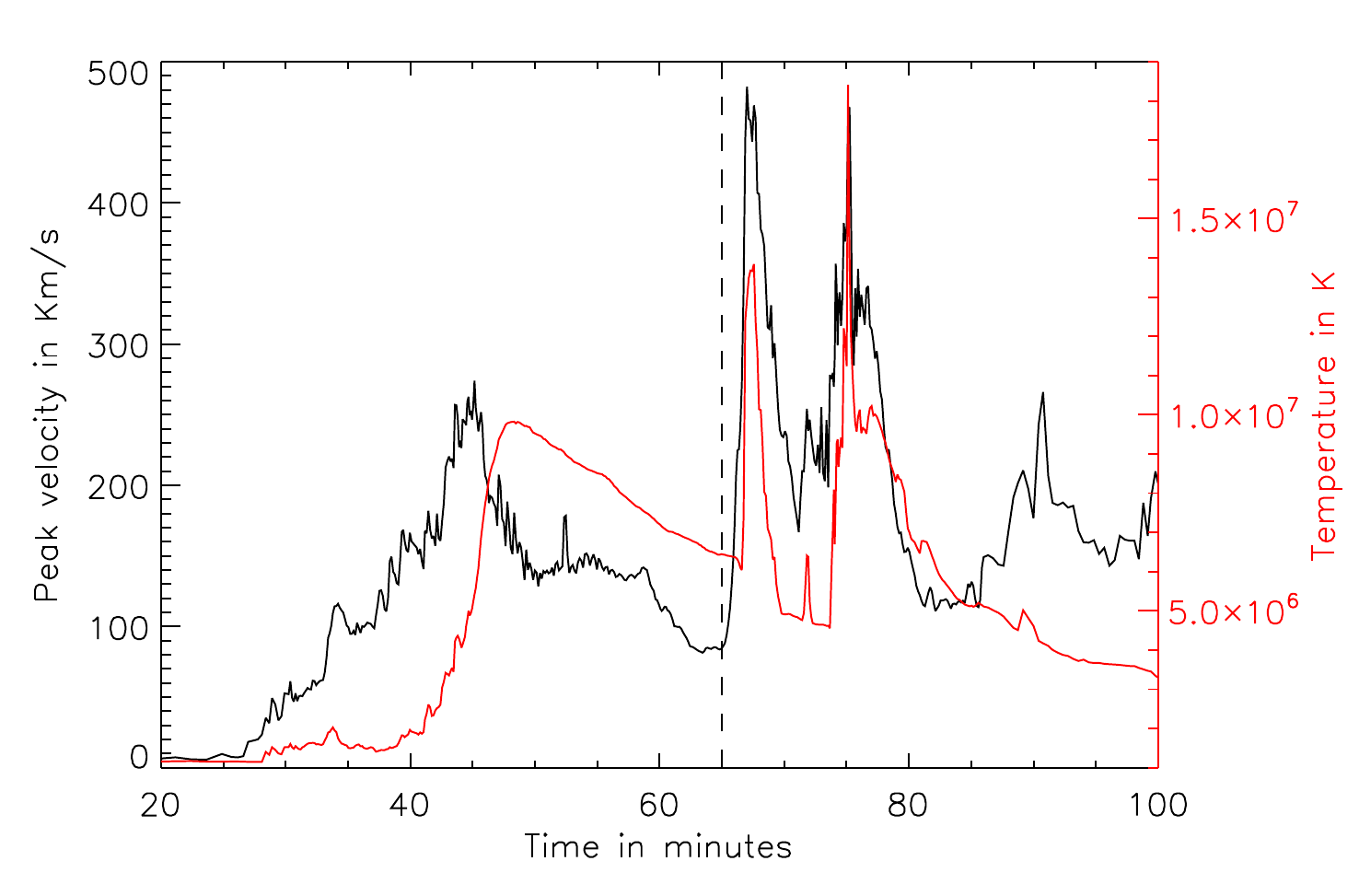}\hfill}
\caption[]{Time evolution of the maximum velocity (black) and temperature
  (red) in a horizontal cross section of the jet at $z=16$
  Mm. To the left of the  
  vertical dashed line one sees the evolution during the main phase of the
  jet. To the right, the violent eruption phase described in
  Sec.~\ref{sec:advanced_phases}}
\label{fig:max_vel_temp}
\end{figure}

If we now turn to the velocities ($|\vvec|$, Fig.~\ref{fig:2Dcut_full},
central panel), the maximum values are not reached in the bulk of the hot and
rarefied material, but, actually, in a thin layer covering the jet as a
blanket.  This can be seen through the horizontal cuts shown in
Fig.~\ref{fig:jet_horizontal_cuts}. The top row shows 2D maps of $|\vvec|$
and $T$ on a horizontal plane cutting the jet at height $16$ Mm above the
photosphere (some 8 Mm above the reconnection region, as indicated by a
horizontal dashed line in Fig.~\ref{fig:2Dcut_full}) at the time of peak jet
activity, $t \sim 45$ min. We clearly see that the jet is actually a curved
sheet-like structure, with plasma moving at velocities around $150$ km
s$^{-1}$ but with a number of individual high-velocity streams embedded in it
(top left panel). The maximum temperatures (top right panel), about $5$
million K, are, again, not uniformly distributed in the sheet, and, in fact,
are not reached at the site of maximum velocity. This situation becomes more
marked along the decay phase of the jet ($t\sim 50$ min, bottom panels): the
high-velocity part of the jet is now (bottom left) a very thin sheet located
on the side of the non-reconnected coronal plasma. On the inner side of that
sheet, a high-temperature domain (a little below $10^7$ K) can still be
found, co-spatial with the low density domain discussed above (bottom right).
The shift between the high velocity jet and the high temperature region seen
in the figure is caused by the decrease in time of the temperature of the
jet: the high temperature region corresponds to older and hotter ejected
material whose motion upward along the jet has been slowed down. A further
feature can be seen in the temperature maps: there is a cool patch
[light-blue, O($10^5$ K)] underlying the hottest blob: this cool patch is not
part of the jet but, rather, corresponds to the topmost reaches of the cool
dense domain described in Sec.~\ref{sec:cold_region}.  The time evolution of
the maximum velocity and temperature in the horizontal plane at $z=16$ Mm is
presented in Fig.~\ref{fig:max_vel_temp}.  During the main phase of the jet
($t<65$ min, i.e., left of the vertical dashed line), the maximum velocity is
about $250$ km s$^{-1}$ and the maximum temperature close to $10^7$ K. The
rise and decay sections of the velocity curve in that phase have a similar
duration, some $15-20$ min in either case; the temperature shows a rapid
rising phase ($5$ min) followed by a more extended decline. After $t > 65$
min, a series of violent eruptions take place, as described in
Sec.~\ref{sec:advanced_phases}.

\subsection{The reconnected coronal loops}\label{sec:hot_loops}

Going back to Figs.~\ref{fig:2Dcut_full} and \ref{fig:3Dview}, we now turn
our attention to the reconnected field lines that are being expelled from the
current sheet toward the right of the figure. Given the magnetic topology,
these are closed magnetic loops (in red in Fig.~\ref{fig:3Dview}), linking
the photospheric domain of the emerged field with the surface roots of the
preexisting coronal field: Fig.~\ref{fig:3Dview} shows that they partially
encircle the emerged dome on one of its sides, constituting an almost
semicircular arcade. They come out of the reconnection layer with a high
level of shear, but turn to configurations nearer to force-free as time
proceeds.  The temperature at the top of the loops follows a pattern as seen
for the jet in Sec.~\ref{sec:the_jet}: first the temperatures are cool,
but then they rise to a peak of order $10^7$ K, which is
reached broadly coinciding with the peak of the red curve in
Fig.~\ref{fig:max_vel_temp}.  Given their high temperatures,
we will often refer to these loops in the following as {\it the hot coronal loops}.

The reconnected loop region grows in size as time proceeds at the expense of
the emerged domain, since new reconnected loops are formed from the latter
that pile up on top of the previous ones.  This can be seen by comparing the
top panel of Fig.~\ref{fig:2Dcut_full} with the corresponding panel of the
earlier time (shown in Fig.~\ref{fig:2Dcut_early}). In fact, we will see in a
later section (Fig.~\ref{fig:wedge}) that the emerged region is being
exhausted  and turns in later phases into a vertically elongated wedge. The
growth of the reconnected loop domain is also related to another phenomenon
which is amenable to observational comparison: the actual jet drifts toward
the positive $x$-direction (i.e., to the right of the 2D plots) as time
advances. This is a consequence of the reconnection process: the
corresponding decrease of the tension of the coronal field allows the current
sheet to move so as to maintain a sufficient stress in the system and
continuously feed the jet with new material.  
Monitoring the sideways motion of the jet on the $y=0$
  plane at $z$ between $10$ and $15$ Mm we obtain velocities of $7$ to $8$ km
\persec\ around $t=30$ min, which then gradually decrease, becoming very low, $2$ km
\persec, by $t=65$ min, i.e., at the time of initiation of the first eruption.

\subsection{The current sheet and the reconnection
  region}\label{sec:current_sheet_reconnection}

\begin{figure}
\centerline{\includegraphics[width=8.5cm]{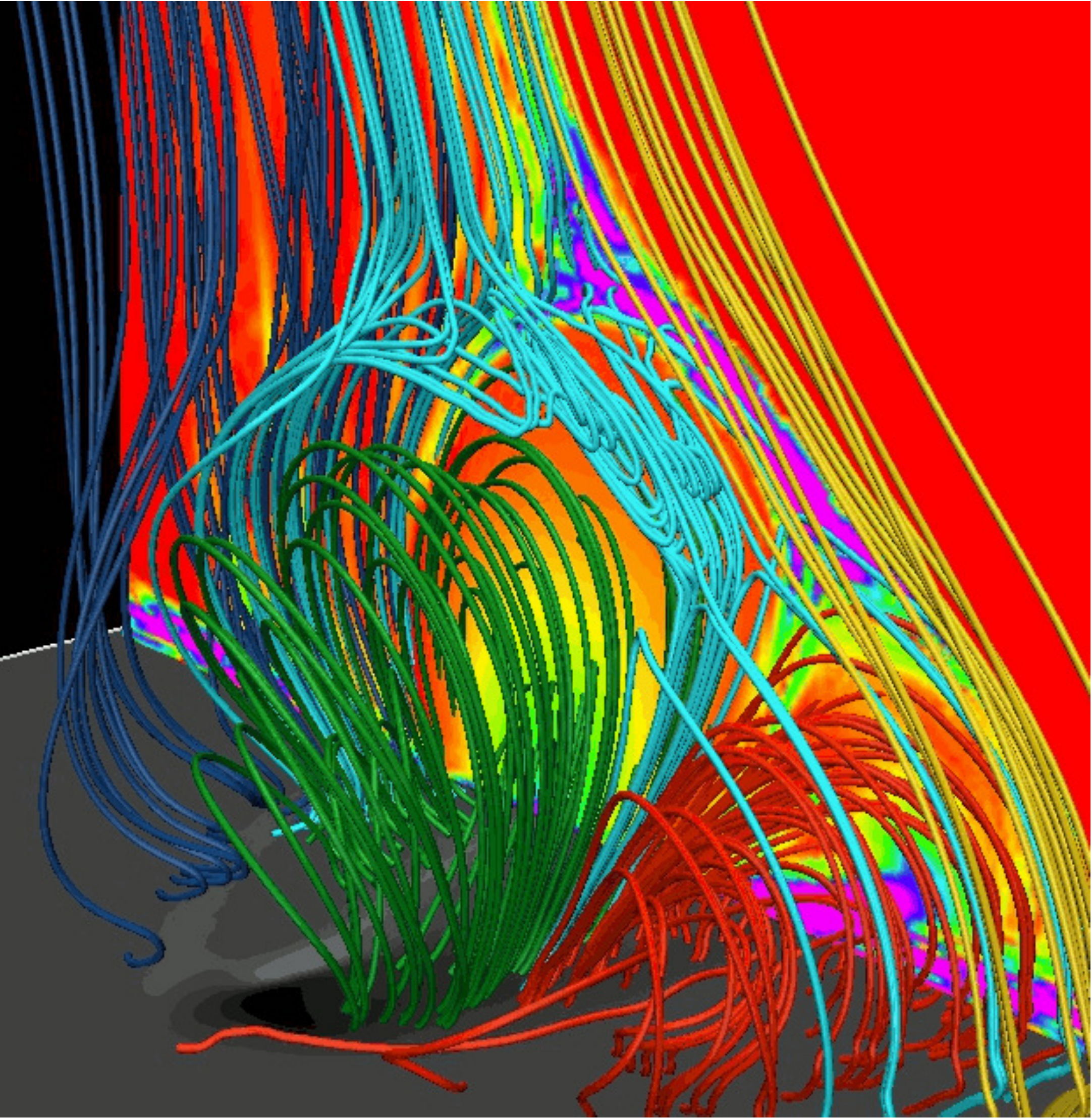}}
\caption[]{Illustration to the field line topology in the current sheet
  explained in Sec.~\ref{sec:current_sheet_reconnection}}
\label{fig:plasmoid_and_null_point}
\end{figure}

The reconnection taking place in the current sheet is of a fully 3D
nature. Going back to the 3D view provided in Fig.~\ref{fig:3Dview}, we can
discern four different sets of magnetic field line systems: (1) at the
top-right, the ambient coronal field lines (yellow) being fed to the
reconnection site joining (2) the emerged magnetic loops (green) which link
the opposite polarities at the surface. On the left, (3) the blue lines are
reconnected field lines that maintain their open-field nature: Even though
they appear similar in shape to the original coronal hole field lines, the
reconnection process and ensuing episodes are causing strong dynamical and
thermodynamical perturbations in the plasma attached to them.  Finally, (4),
at the bottom right, a set of closed magnetic loops (red) as described in
Sec.~\ref{sec:hot_loops} that link the photospheric domain of emergence with
the external photospheric field.
The magnetic topology shown in Fig.~\ref{fig:3Dview} looks deceivingly
simple. On the one hand, there appear plasmoids within the current sheet,
which, given their 3D nature, are in fact solenoids wrapping around the
general hill-shape of the emerged material
\citep[see][]{2006ApJ...645L.161A}.  This is shown in
Fig.~\ref{fig:plasmoid_and_null_point}. The vertical panel contains a map of
$|\jvec|/|\Bvec|$: the purple-colored concentrated current sheet is
apparent. The light-blue field lines were traced so that they pass
preferentially near the null points.  The other field line sets in the figure
are as described in Fig.~\ref{fig:3Dview}. The light-blue lines clearly show
a collection of 3D plasmoids, with a solenoidal appearance, embedded in the
collapsed current sheet. In fact, additionally to using visualization
techniques, we could detect the presence of a number of null points within
the current sheet using a null-point finder program
\citep{2007PhPl...14h2107H}; they turn out to be contained in a small range
of the axial ($y$) coordinate, and slowly move in the direction of the
negative surface polarity as the jet develops. The presence of plasmoids
indicates that multiple episodes of tearing instability have occurred in the
sheet.

\begin{figure}[h]
\sizefig=12.cm \ifnum \value{twocol} > 0 \sizefig=9cm \fi
\centerline{\hfill
\includegraphics[width=\sizefig]{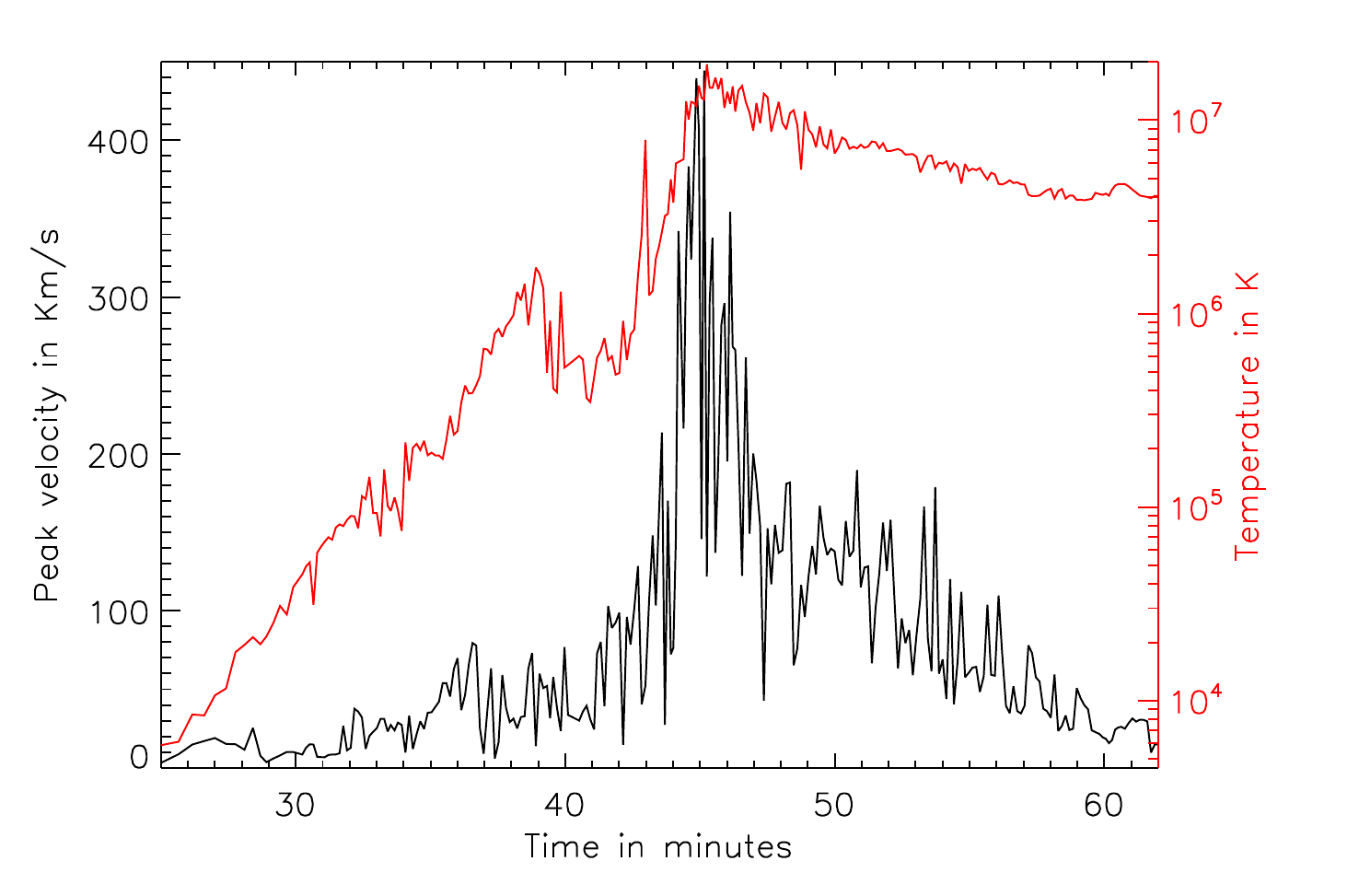}\hfill}
\caption[]{Time evolution of velocity (black) and temperature (red) 
  in the current sheet, tracked by determining the location of minimum $|B|$.
  When comparing with Fig.~\ref{fig:max_vel_temp}, note the linear
  temperature scale in that figure.}
\label{fig:time_evolution_in_current_sheet}
\end{figure}

The time evolution of the current sheet can be measured through various
quantities. As a proxy for the location of the reconnection site we have
chosen the point in the current sheet, $\rvec_{rp} \equiv (x_{rp}, y_{rp},
z_{rp})$, where $|\Bvec|$ reaches its minimum, $B_{min}$. This procedure does
not lead to an accurate position determination, but is good enough to provide
approximate numbers for the temperatures and velocities near the reconnection
site, probably in the neighborhood of a null point.
In Fig.~\ref{fig:time_evolution_in_current_sheet}, the temperature (red) and
velocity (black) measured at $\rvec_{rp}$ are plotted for the duration of the
main phase of the jet.  We can see that the rise and decay of the temperature
and velocity in the current sheet correlate well with those in the jet
discussed in Sec.~\ref{sec:the_jet} (see Fig.~\ref{fig:max_vel_temp}). The
peak velocity, in particular, is higher than the corresponding peak value
in Fig.~\ref{fig:max_vel_temp}. By tracking in time the
  position of $B_{min}$, we see that it rises by $\sim 5$ Mm before remaining
  at $z=7$ Mm in the decay phase; horizontally, it also slides by $\sim 7$ Mm
  into the $y>0$ half space (where the negative photospheric polarity is
  located). Instead, it does not move much in $x$ (i.e., the abscissas of
  Figs.~\ref{fig:2Dcut_early} or \ref{fig:2Dcut_full}): it looks as though
  the region with the null points is {\it sliding around the flanks of the emerged
    mountain}. This impression is confirmed by following the actual motion of
  the null points in time and will be further discussed in Sec.~\ref{sec:advanced_phases}.

\section{The cold periphery: the dense wall}
\label{sec:cold_region}

\begin{figure*}
\centerline{\includegraphics[width=16cm]{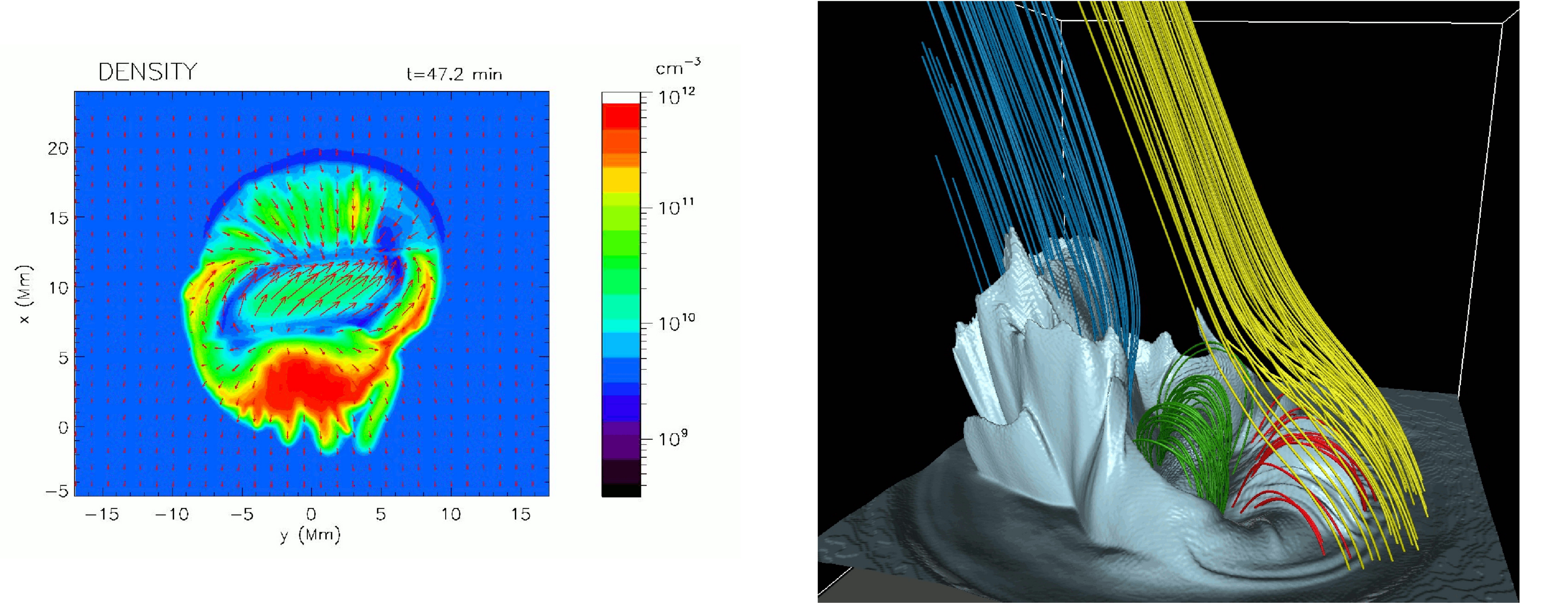}}
\caption[]{Illustration of the high-density wall. Left: 
density map on a horizontal plane $4$ Mm above the photosphere -- the arrows
show the horizontal projection of the magnetic field vector. Right: 3D
visualization of the dense domain using a density isosurface at $\rho=4\,10^{10}$
\percc\ and field lines from the different magnetic connectivities illustrated
in Fig.~\ref{fig:3Dview}
} 
\label{fig:density_wall_hor_cut}
\end{figure*}

In previous sections we mentioned the presence of a relatively cold and dense
volume surrounding the emerged magnetic loops as a wall.  A visualization of
this structure is provided in Fig.~\ref{fig:density_wall_hor_cut}.  The panel
on the left contains a density map on a horizontal cut at $z=4$ Mm and $t=47$
min, i.e., near the peak phase of the jet activity. The map clearly shows the
very dense surroundings of the emerged-loop region, like a ring or wall, with
densities generally between $10^{11}$ and $10^{12}$ \percc, peaking at the
back of the structure, i.e., below the jet. In the figure, arrows
corresponding to the horizontal projection of the magnetic field have been
added. The location of this ring or wall with respect to the emerged loop
system can be seen in the 3D image in the right panel of the figure: the
image contains an isosurface of the number density for $n = 4\, 10^{10}$
\percc\ (bright to dark grey)  and includes selected
field lines of the different magnetic systems (same color code as for
Fig.~\ref{fig:3Dview}). The isosurface has a 
crater-shape with ragged peaks around a hole, and is continued sideways by a
roughly horizontal domain until the boundaries of the box. In the central
hole we see the (green) magnetic arcade of the emerged domain.  The
reconnected coronal loops (red) are almost covered by this isosurface; the
highest density peaks are located in the reconnected region below the jet
(blue field lines). Both reconnected systems have inherited the high
densities of the emerged material. At this time, however, the emerged region
is already very rarefied (Sec.~\ref{sec:the_jet}, Fig.~\ref{fig:2Dcut_full},
bottom panel), and this is the reason for the central hole of the density
isosurface. The maximum elevation of the isosurface is several Mm. The dense
wall is clearly a 3D phenomenon: it appears all around the emerged domain,
although its shape depends on the actual location relative to the emerged
flux system. The dense domain has transition region
  temperatures, between a few times $10^4$ and a few times $10^5$ K, well
below coronal values.

\begin{figure*}
\centerline{\includegraphics[width=18.cm]{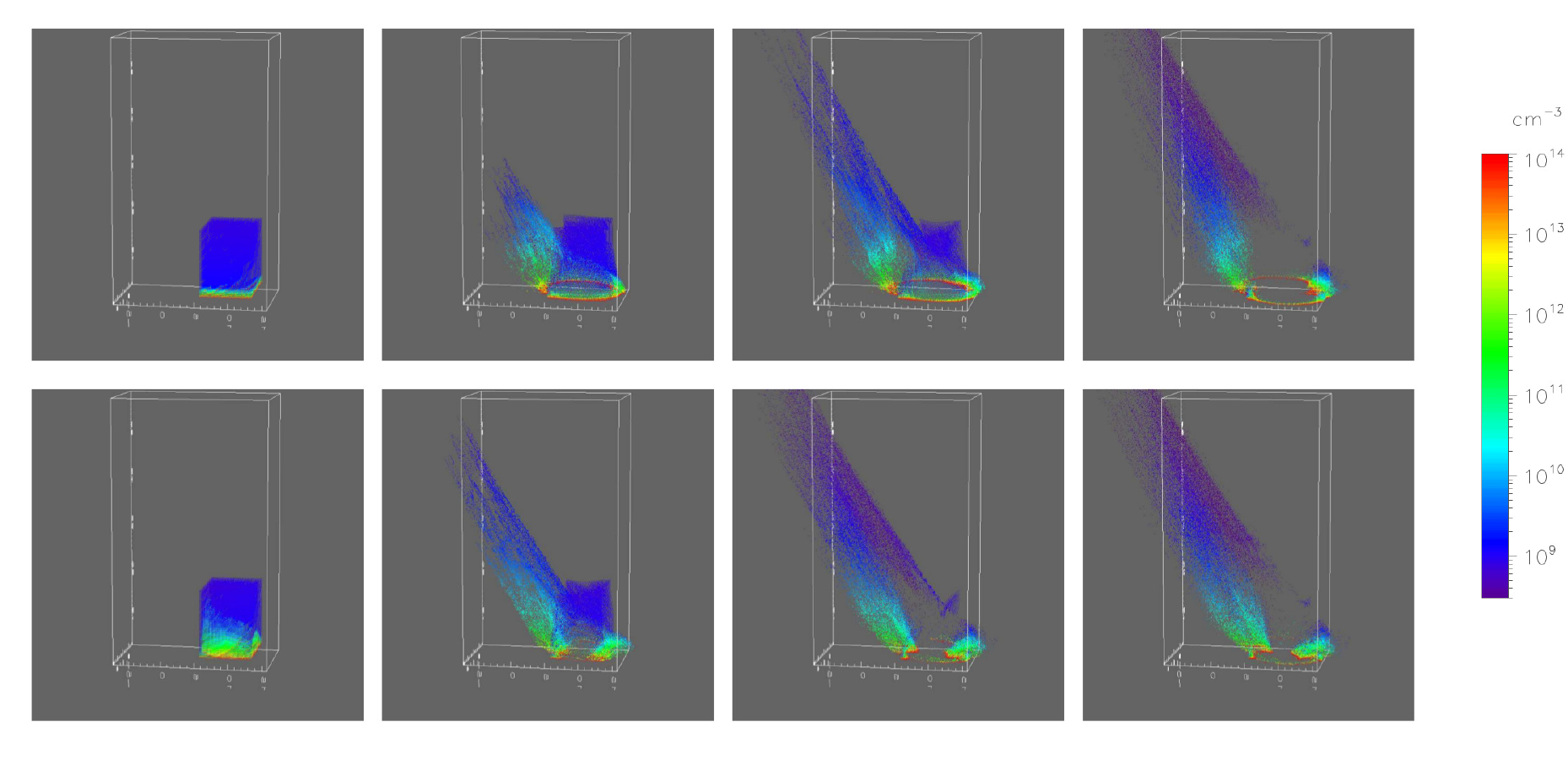}} 
\caption[]{\label{trace_particles.fig} The frames show the time evolution of
  the Lagrange tracing particles. Each row corresponds to a different tracing
  sequence, with the leftmost frame being the respective initial
  condition. The times for the top row frames are $t=20.8$, $35.4$, $43.8$
  and $62.5$ min; those for the bottom row are $t=45.8$, $51.7$, $57.5$ and
  $62.9$ min. The color coding for the particles corresponds to the local plasma
  number density. See also the accompanying movies.}
\end{figure*}

To disclose the origin and nature of this dense wall, we have used Lagrange
tracing for the plasma elements. For the tracing, we interpolate $\vvec$
between snapshots, which had been taken with high time cadence (between $8$
and $16$ sec in this phase).  Figure~\ref{trace_particles.fig} shows the
position of the tracer particles at different times in the experiment. For
the figure, we selected an initial box with tracers distributed uniformly
between approximately $z=-0.3$ Mm and $z=20$ Mm, and covering horizontally a
large fraction of the emerging region. The top of the box is located
therefore some $10$ Mm above the top of the emerged dome in the main jet
phase.  Each particle is given a color according to its instantaneous density
(colorbar on the right). We note that green ($n \sim 6\,10^{10}$ to
$2\,10^{12}$ \percc, say) corresponds to heights $1.9$ to $2.8$ Mm in the
initial atmosphere, whereas dark blue ($n \sim 3\ 10^8$ -- $10^9$ \percc)
corresponds to coronal densities.  We had two different Lagrange-tracing runs
(top and bottom rows in the figure, respectively), with initial boxes of the
same size and position and with the same initial tracer distribution, but
with different starting times.  The sequence in the top row was started at
time $t=20.8$ min (left panel), i.e., when the top of the rising flux tube
was still at the photosphere. The tracers in the initial Lagrange box with
green and blue colors thus correspond to plasma that was in the atmosphere
before the flux emergence takes place, attached to inclined coronal field
lines. The times for the other three panels in that row are, from left to
right, $35.4$, $43.8$ and $62.5$ min, respectively.

The panels in the top row show how the mass elements in the atmosphere change
position following the arrival of the tube. At the time of the second and
third panels, the magnetic tube has already fully emerged (its top is in fact
located at about $z=8$ Mm). Most of the tracer particles of the pre-emergence
material have been moved to the left where they constitute part of the dense
region visible in Fig.~\ref{fig:density_wall_hor_cut} (right panel, e.g., the
volume delineated by the grey isosurface). The way in which that happens is:
those particles are attached to coronal field lines that somewhere along
their length are pulled down into the approaching reconnection region. This
of course does not imply that the Lagrange particles themselves go through
the reconnection domain: it is the field lines they are attached to that must
go through the (spatially small) diffusion region where the change of
magnetic connectivity takes place. Those that reach the region on the left at
early stages end up in the dense-volume visible on the left hand side. The
two rightmost panels show how the particles that go over to the reconnected
side in the more advanced stages are likely to be launched along the jet with
high speeds (keeping a density on the order of their initial coronal value --
see the deep-purple set of particles in the upper half of the box). The
rightmost panel also shows how at that late time the pre-emergence material
has been completely vacated from the position of the emerging dome.  The
velocities found in the dense domain on the left are not large ($\ \lesssim 50$
km \persec).

The lower row of panels illustrates further aspects of the plasma
evolution.  This sequence was started at time $45.8$ min of the
simulation, i.e., near the third panel from the left in the top row. A
large part of the initial Lagrange box for this series contains
emerged material. We see in the second and third panels (green colored
particle domains) how a fraction of these particles go over to the
dense wall domain. Those particles are attached to field lines that
are undergoing reconnection.  As explained for the top row, however,
the particles themselves need not go through or near the reconnection
point: their field lines take the particles with them over to the
reconnected domain, without necessarily being part of the
high-velocity reconnection outflows or jets. Analyzing the images in
this series, one could say that the emerging dome is progressively
peeling itself off of its external layers, passing its dense external
layers little by little onto the reconnected systems (open field,
closed-loop systems). Again here, the velocities for the green-colored
particles in the third rightmost panels in this row are not large
($\ \lesssim 50$ km \persec). On the other hand, another fraction of
the particles really go near the reconnection site, get accelerated
and end up being part of the jet, hurled upward with the high speeds
mentioned in Sec.~\ref{sec:jet_reconnection}. As a final item, a
fraction of the particles in either time sequence are attached to
field lines that end up in the reconnected loop system on the right of
the figure.

The formation of a dense circular wall constituted in part by debris
from the emerged material seems to be a rather natural feature in the process
described in this paper. One wonders if any part of the cool and dense
material discussed so far could really be seen as constituting a cool
H$\alpha$ surge or CaII jet, as proposed in the literature \citep[see,
  e.g.,][]{1996PASJ...48..353Y, 2003ApJ...593L.133M,
  nishizuka_etal_2008}. This is discussed in Sec.~\ref{sec:discussion}.

\section{Advanced phases}\label{sec:advanced_phases}

The  rather smooth evolution described in the previous sections
ends abruptly at  $t\sim 65$ min. There follows a more dramatic
phase, with several violent eruptions occurring in different parts of the
emerged structure: to the end of the experiment ($t=100$ min), we have
located five such eruptions, and there is no reason to assume the
series would stop at that time. In this section we 
discuss these events and their relation to eruptions found in previous
flux emergence experiments.

\subsection{First eruption: a variation of the sheared arcade instability}
\label{sec:first_eruption}

The field line configuration right before the start of the first
eruption is illustrated in Fig.\ref{fig:wedge}: in the top panel, a
magnetogram, i.e., the grey-scale map of $B_z(z=0)$, shows the bipolar
region with its two strong opposite polarities which is clearly
developing a sigmoidal shape of positive helicity.  Two sets of field
lines have been drawn in that panel: the one in red corresponds to the
inner core of the emerged flux rope, which is not far above the
photosphere and very nearly delineates the polarity inversion line
(PIL).  The green field lines identify the magnetic system that is
still remaining in the emerged plasma domain. To better illustrate
this, the central panel additionally shows a map of the electric
current (precisely: of $\joverB$) on the vertical plane at $y=0$. We
see in the image that the green field lines of the previous panel go
through the cross section of what remains of the emerged dome: the
continued peeling off of plasma and field lines from the latter has
turned it into a thin wedge with a narrow base. We also recognize a
strong and thin current sheet (purple) toward the upper-right corner,
topping up the wedge.  To the right of the wedge in the figure, the
hot-coronal loop region (which was a relatively small feature in the
early stages, see Figs.~\ref{fig:2Dcut_full} and \ref{fig:3Dview}) has
become quite large: we have drawn a bunch of field lines (in light
brown), which show the general geometry of the field there.  We see
that a thin, almost vertical current sheet has built up at the base of the
wedge; to 
investigate its nature, in the bottom panel we draw in blue a further
set of field lines (and remove the green and brown ones for clarity):
the new set has been selected so that they pass through the wedge cross
section at intermediate heights; we see that those lines are part of a
highly sheared magnetic arcade. We also see the reason for the current
concentration: the arcade has footpoints of opposite polarity near
each other, ready to undergo reconnection.

\ifnum \value{twocol} > 0
\begin{figure}
\else
\begin{figure*}
\fi
%
\sizefig=6.8cm \ifnum \value{twocol} > 0 \sizefig=7.5cm \fi
\vbox{
\centerline{\includegraphics[width=\sizefig]{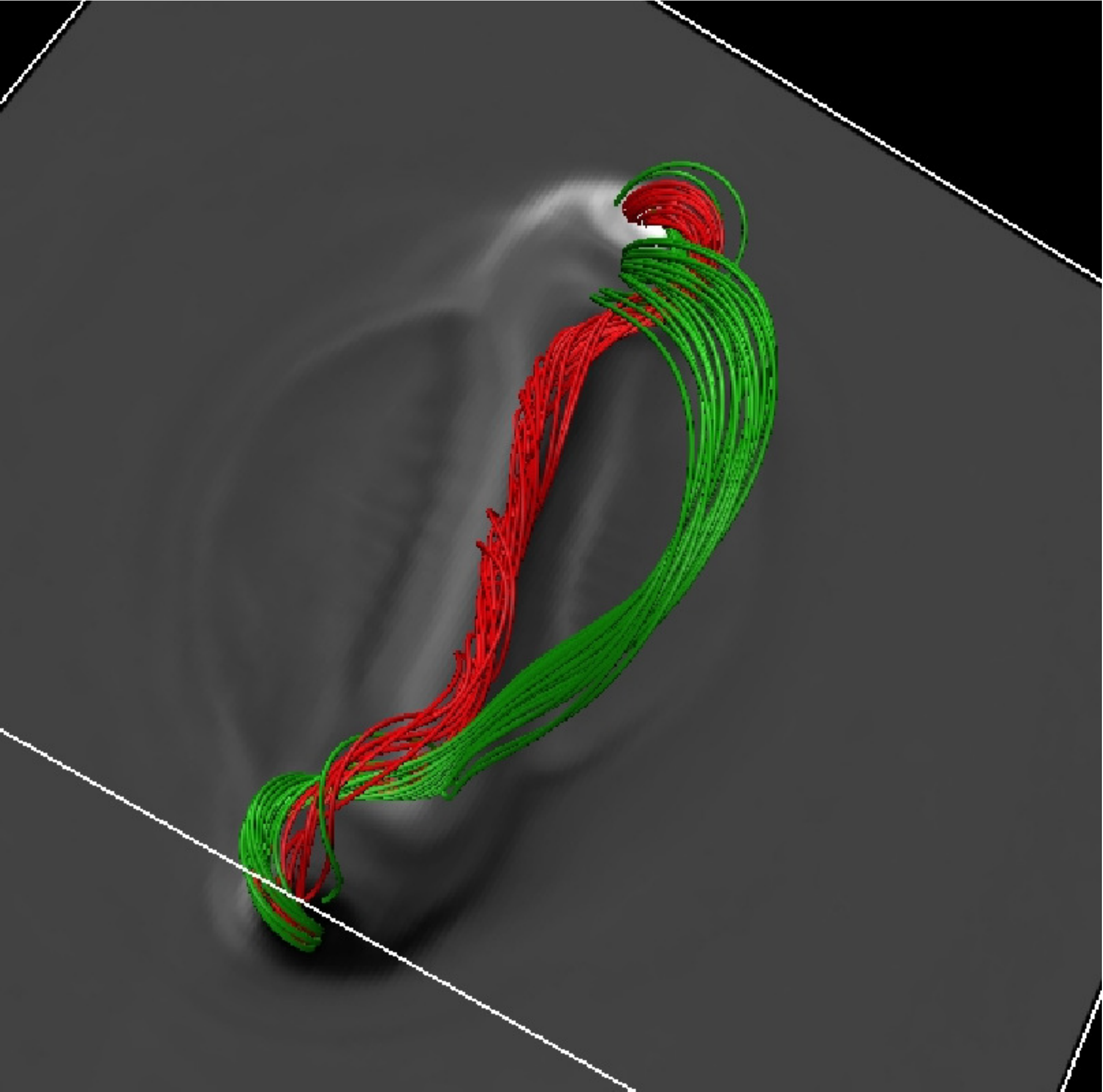}}
\vspace{2mm}
\centerline{\includegraphics[width=\sizefig]{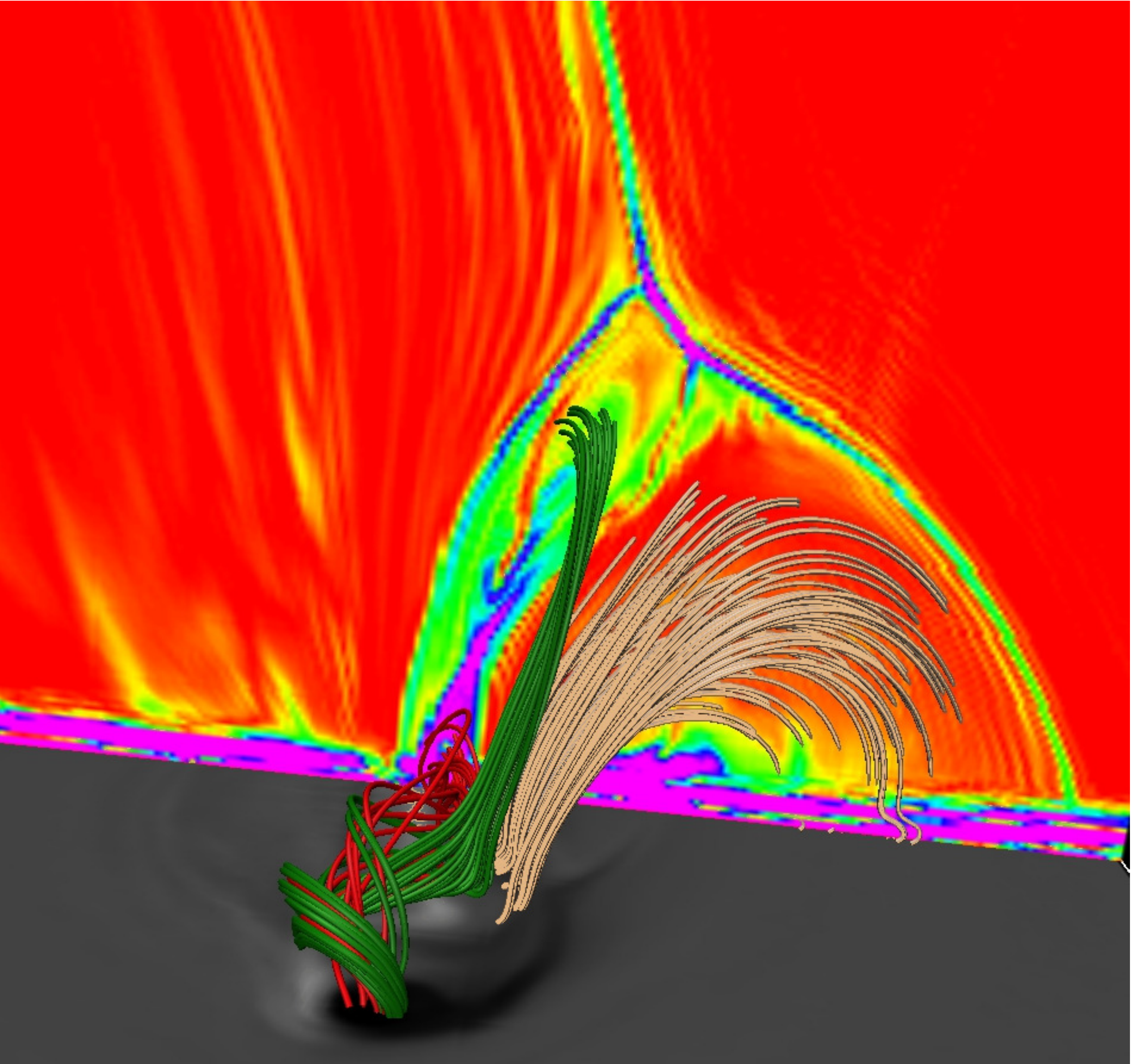} }
\vspace{2mm}
\centerline{\includegraphics[width=\sizefig]{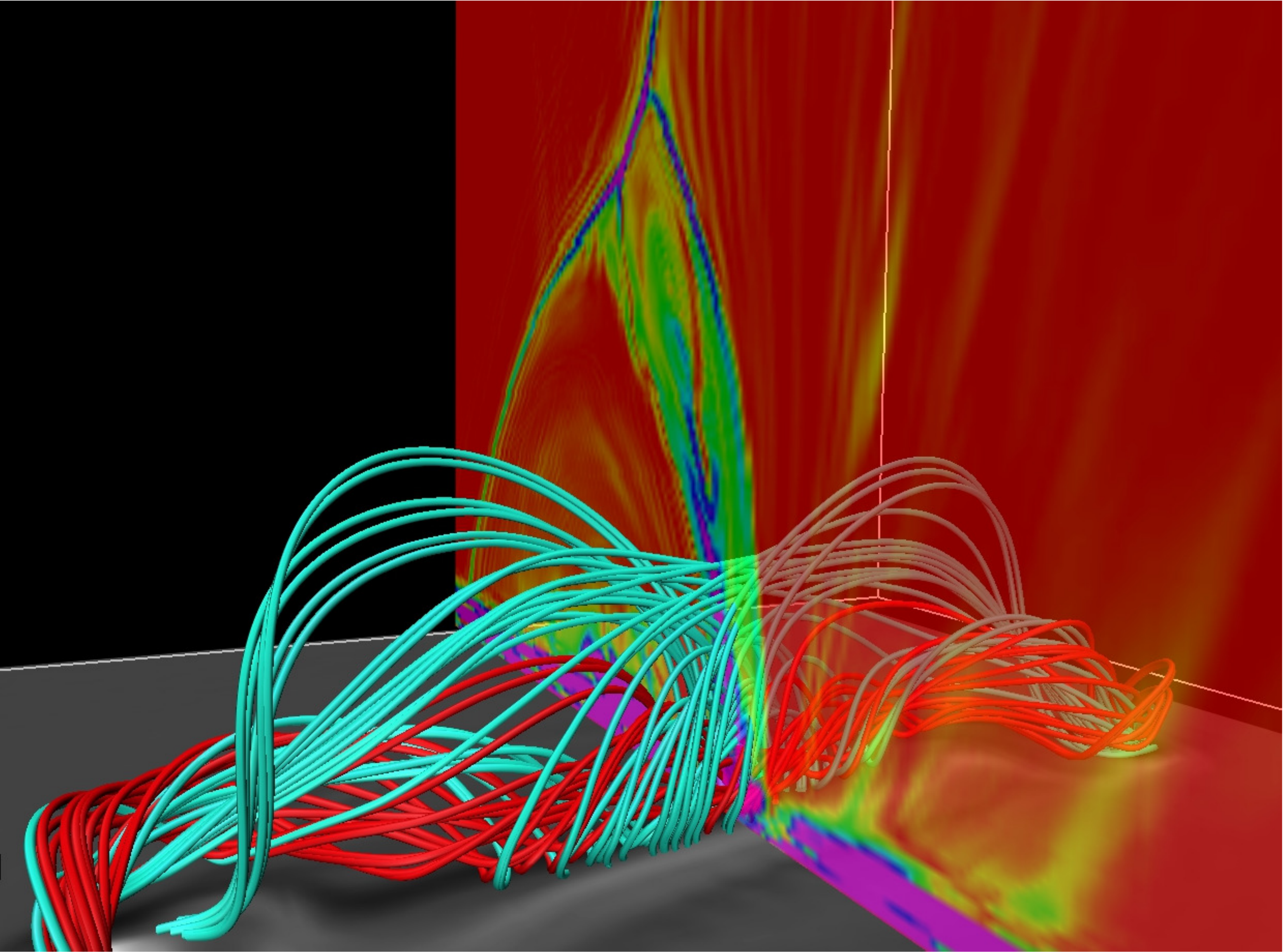} }}
\caption[]{Illustration of the field-line and electric current
configuration right before the start of the first eruption ($t=64.5$
min). The grey-scale map in all panels is a {\it magnetogram}, meaning the
distribution of $B_z^{}$ at the level $z=0$.
The vertical color map (central and bottom  panels) corresponds
to $\joverB$ in arbitrary units, and 
extends between red (lowest value) and purple (highest value). The different
field line sets illustrate characteristic connectivity patterns, 
as explained in the text. Note that the vertical plane
  is seen from opposite sides in the middle and bottom panels.}
\label{fig:wedge} 
\ifnum \value{twocol} > 0
\end{figure}
\else
\end{figure*}
\fi

\begin{figure}
\sizefig=9.cm \ifnum \value{twocol} > 0 \sizefig=8cm \fi
\vbox{
\vspace{3mm}
%
\centerline{\includegraphics[width=\sizefig]{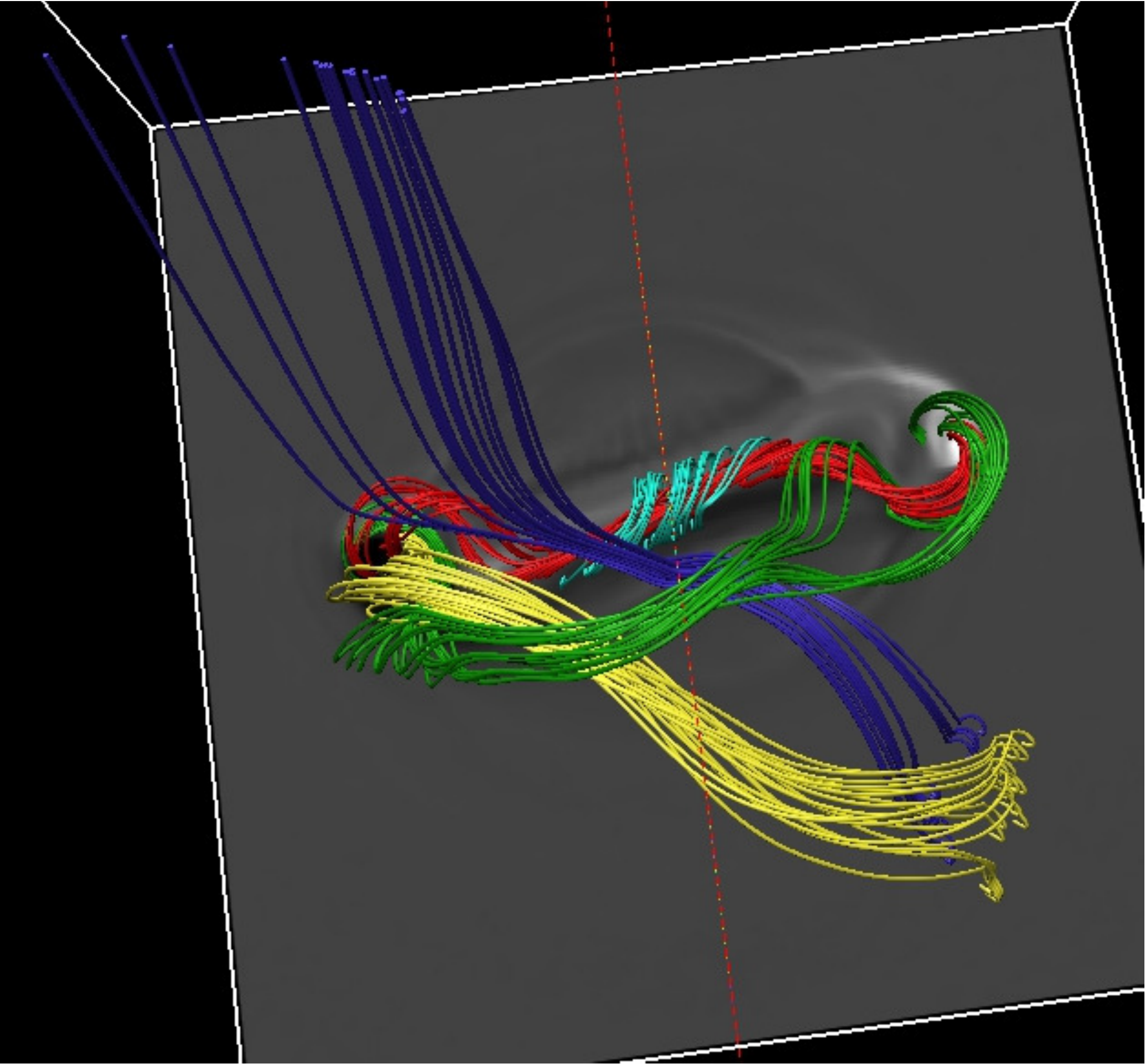}}
\vspace{3mm}
\centerline{\includegraphics[width=\sizefig]{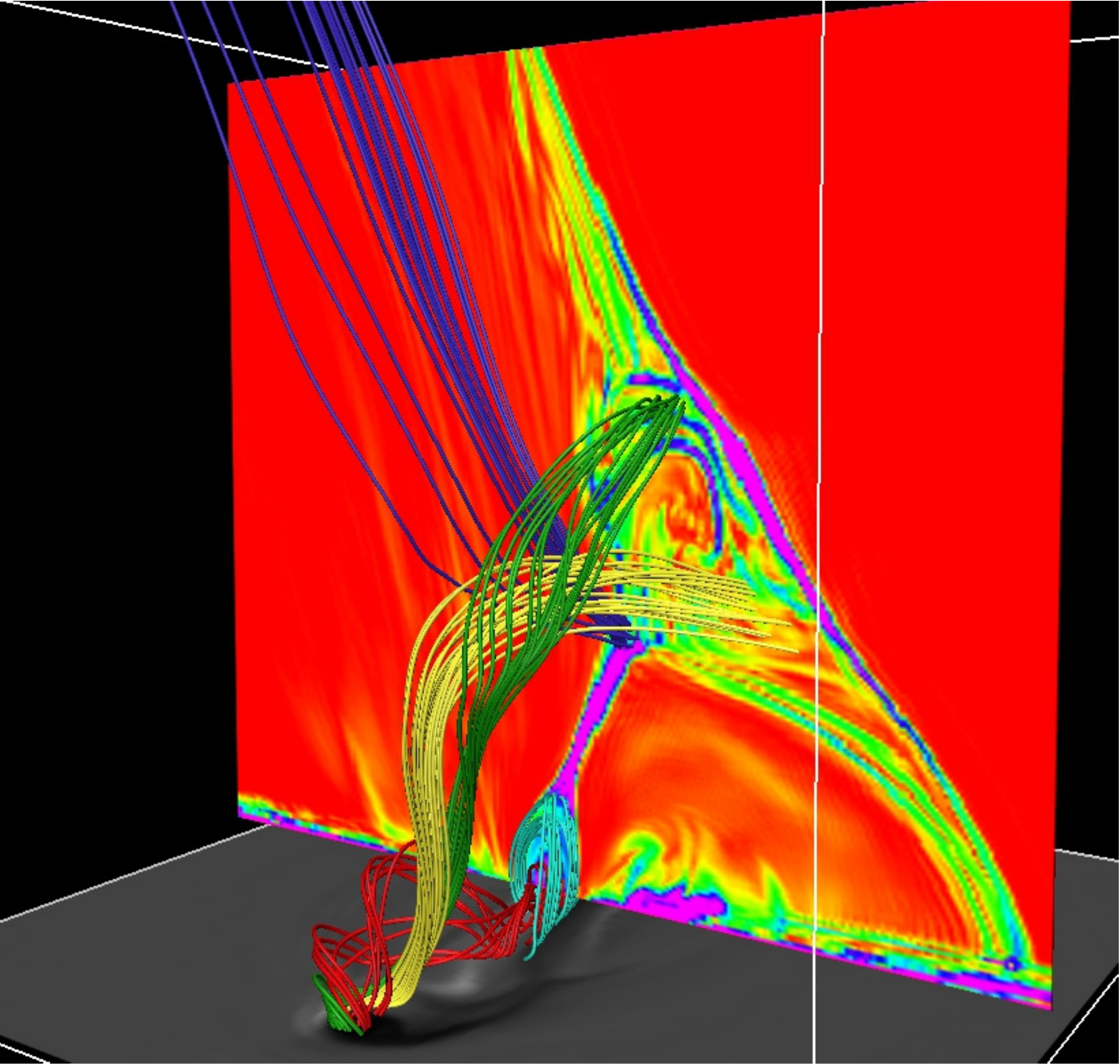}}}
\caption[]{Field line connectivity and current distribution as the
  first eruption is unfolding $t \sim 65.9$ min.  Upper panel: top
  view showing the magnetogram at the $z=0$ level and five sets of
  field lines representative of major connectivity patterns at that
  time. Lower panel: map of $\joverB$ on the central vertical plane
  ($y=0$), showing the current sheet located underneath the flux rope
  that contains what remains of the emerged material.}
\label{fig:eruption_1_a_2}
\end{figure}

The first eruption takes place in the time frame between,
approximately, $t=64$ minutes and $t=70$ minutes. The
opposite-polarity feet of the field lines in the sheared arcade (blue
field lines in Fig.~\ref{fig:wedge}, bottom panel) get close enough to
each other for fast magnetic reconnection to start. The first stages
of this process strongly resemble the {\it tether cutting}
reconnection scheme extensively discussed in the literature
\citep[see][]{1988ApJ...328..830M,1989SoPh..121..387S,1992LNP...399...69M,
  2010ApJ...720..757M}, in particular in relation with the emergence
of flux ropes through the photosphere \citep{2004ApJ...610..588M,
  2008A&A...492L..35A, 2008ApJ...674L.113A, 2012A&A...537A..62A}: the
reconnection creates two sets of field lines at either end of the thin
quasi-vertical current sheet created at the feet of the sheared
arcade; the one at the top is a twisted magnetic rope that is hurled
upward whereas the one below is a simple arcade that progressively
turns into a quasi-potential configuration, as described below. The
reconnection process, however, soon becomes much more complicated than
in the standard flare situation. The reconnection between opposite
polarities at the feet of the original sheared arcade is asymmetric:
the arcade field lines on one of the sides of the current sheet become
fully reconnected while there is still remnant arcade flux on the
other side. As a result, field lines from the hot loop system (like
some of those in light brown in the central panel of
Fig.~\ref{fig:wedge}) start to be reconnected with those remnant field
lines from the arcade (light-blue field lines in
Fig.~\ref{fig:wedge}). In the final stages of the eruption, field from
the open magnetic system (not shown) at the far end of the wedge in
Fig.~\ref{fig:wedge} get in contact with field lines from the hot loop
system (light brown) and reconnect across the current sheet.  The
resulting connectivity scheme can be discerned through
Fig.~\ref{fig:eruption_1_a_2}. In the top panel we are showing bunches
of field lines illustrating five of the different connectivity
patterns just discussed: the green field lines are the heirs of the
green field line system in Fig.~\ref{fig:wedge} and are ejected toward
the beginning of the reconnection process; the yellow and dark blue
field line sets correspond to the new connectivities created in the
advanced stages of the eruption, namely sheared-arcade connected with
hot coronal loops (yellow) and open field lines connected with hot
coronal loops (dark blue). The other two sets are best discussed using
the bottom panel of the figure: in it we are including a color map of
$\joverB$ on the vertical plane at $y=0$ that clearly shows (a) the
thin current sheet below the erupted material (in purple), (b) the
current sheet at the interface between the erupted material and the
ambient coronal field, and (c) the complication of the erupted region,
in which plasma attached to the field lines of the different stages of
the reconnection is moving with high speed and expanding sideways. In
this panel we are also showing a small arch system
near the photosphere (in light-blue color, but
not to be confused with the light-blue system in
Fig.~\ref{fig:wedge}): this is indeed the result of reconnection at
the bottom end of the thin current sheet; we note that it has a much
lower shear than the original arcade. The red field line system,
finally, is the original core of the emerged flux rope, essentially
the same system as shown in red in Fig.~\ref{fig:wedge}.

The erupting material in the successive stages is hurled onto the overlying
open field with high velocity, around $600$ km s$^{-1}$ at its core.  The
ambient coronal magnetic field outside of the jet is strong enough to
withstand the impact; the eruption, therefore, is channeled along the field
lines where the original jet was propagating and disturbs their
neighborhood. The eruption will be seen not only to perturb a large part of
the emerged domain (as explained below), but, also, to propagate as a sort of
short-lived, impulsive jet with a complex structure roughly in the same
direction of the original jet.  The differences in field line orientation
between the erupted material and the open field create the thin current sheet
that delimits the perturbed domain toward the top-right corner in the lower
panel of Fig.~\ref{fig:eruption_1_a_2}, through which reconnection occurs.
The strong perturbation caused by this eruption is communicated along the
field lines and ejected upward reaching the top levels of the numerical box.
The maximum temperature and velocity in the horizontal
  plane $z=16$ Mm plotted in see Fig.~\ref{fig:max_vel_temp} show a
  pronounced peak of about $1.4\,10^7$ K and about $500$ km \persec ,
  respectively, during this eruption. The kinetic energy in the volume where
the eruption takes place goes through a marked peak of $8\,10^{26}$ erg.

\subsection{The off-center eruptions}\label{sec:later_eruptions} 

Four further eruptions occur in the remaining $\approx 30$ minutes of the
experiment: they all start and mainly take place in the
{\it flanks} of the active region, i.e., in the general volume above the
strong polarities at the photosphere. Their evolution also has a
number of further features that distinguish them from the first eruption, and
also among each other. In the following we describe the main traits of the
first two of these off-center eruptions: covering all of them 
goes beyond the scope of the present paper.

The eruption discussed in the previous subsection
(Sec.~\ref{sec:first_eruption}) starts in the low coronal heights midway
between the two opposite polarities at the photosphere (i.e, around
$y=0$). Yet, the large perturbation quickly gets communicated to the flanks,
i.e., to the levels above the strong photospheric polarities. This occurs
through a variety of phenomena: referring to Figs.~\ref{fig:wedge} and
\ref{fig:eruption_1_a_2}, first, not only does the erupting flux rope (set of
green field lines in either figure) rise, but, also, its roots reach
regions nearer the two opposite photospheric polarities; second, the number
of hybrid, reconnected field lines of the kind colored in yellow in
Fig.~\ref{fig:eruption_1_a_2} increases as time goes on, thus communicating
the perturbation to the regions nearer the negative polarity visible in that
figure (i.e., toward the $y>0$ half-space). However, there is a region
roughly above the PIL close to the strong negative photospheric polarity that
is not very much affected by the first eruption: this coincides with the
strongly curved, hook-like shaped region of the PIL near the negative
polarity at the surface apparent in Figs.~\ref{fig:wedge} (top panel) and
\ref{fig:eruption_1_a_2}. The unperturbed region can also be seen
in Fig.~\ref{fig:preeruption2}, top panel, where a vertical map of $\joverB$
in the neighborhood of the strong negative polarity is shown corresponding to
$t=66$ min (same time as for Fig.~\ref{fig:eruption_1_a_2}). Three separate
closed regions (labeled regions 1 through 3) can be discerned adjacent to
each other in the plane, two with a wedge shape and a roundish one on the
right: the central one (region 2) is the result of the strong perturbation
caused by the first eruption; the upward-pointing wedge-shaped region to the
left (region 1) contains magnetic field of the initial emerged tube: two
field line sets have been drawn: the bottom one (in red) is quite twisted and
considerably inclined; it nearly follows the curved, hook-like shape of the
PIL near the negative surface polarity; the field-line set in green is less
twisted but, like the red one, links to the regions near the PIL on the side
of the vertical plane of the figure. These field line sets appear not to have
changed topology nor their general shape as a result of the eruption at the
center: we will see presently (Sec.~\ref{sec:second_eruption}) that a violent
eruption takes place shortly after the time of the figure starting from field
lines like the red ones in that panel.  The roundish area on the right
(region 3), finally, corresponds to the hot coronal-loop region which, again,
is not greatly affected by the eruption. In the following we describe two
events that take place in this end (the $y>0$ end) of the emerged domain:
namely, the reconnection event occurring near the prominent null point
situated in that neighborhood (Sec.~\ref{sec:null_above_negative_polarity}),
and the development of the eruption just mentioned.

\subsubsection{The null point above the negative photospheric polarity}
\label{sec:null_above_negative_polarity}

The complicated null-point and plasmoid structure of the
  reconnection site described in Sec.~\ref{sec:current_sheet_reconnection}
  (Fig.~\ref{fig:plasmoid_and_null_point}) becomes increasingly simple as it
  slides sideways along the $y$-direction in the decay phase of the jet
  toward the negative-polarity end of the active region; at time $t=50$ --
  $55$ min, in fact, the magnetic skeleton here contains a single null point
  with real eigenvalues. At this time the fan surface is locally near
  horizontal; on the one side it represents the canopy of the hot loop system
  and, on the other, it divides the emerging flux from the open flux
  region. The electric current in the fan surface near the null increases its
  intensity in the phase up to the second eruption; doing this the current
  sheet expands in width, and the distance between the {\it roots} of the two
  spine axes move apart, attaching to the edges of the current sheet (See
  \citealt{2007PhPl...14e2109P} for a detailed description of this kind of
  null collapse).

The location of the null and the structure of its skeleton at the time
of the first eruption ($t=66$ min) is illustrated through the light
green and red sets of field lines in the central panel of
Fig.~\ref{fig:preeruption2}. A plasmoid (delineated by the green field
lines as labeled in the panel) is seen to occupy the site of the
current sheet, indicating that the latter has undergone a
tearing instability.  This behavior of null point evolution is seen in
various dedicated numerical experiments of asymmetric single 3D nulls
that are being stressed \citep{2011A&A...534A...2G}. The reconnection
process that we can deduce from the numerical data is of the type
known as spine-fan reconnection \citep{2009PhPl...16l2101P}, where
flux is being transported through the fan surface and spine axis from
one topological domain to another.  In fact, as illustrated in
Fig.~\ref{fig:preeruption2} (central panel), the reconnection involves the flux rope
of the previous eruption (red field lines in the figure): as it
continues to reach higher levels, it starts interacting with the
stressed null point and the field line connectivity changes, allowing
these field lines to connect the positive surface flux concentration
with the open field region, whereas, in turn, the left-hand side
footpoint becomes connected to field lines near the downward-pointing
spine.

The comparatively simple topological structure just
  described is deeply modified by the second eruption, which causes the
  creation of a cluster of null points in the current sheet.

\ifnum \value{twocol} > 0
\begin{figure}
\else
\begin{figure*}
\fi
\sizefig=6.0cm \ifnum \value{twocol} > 0 \sizefig=6.5cm \fi
\centerline{\hfill
\vbox{
\centerline{\includegraphics[width=\sizefig]{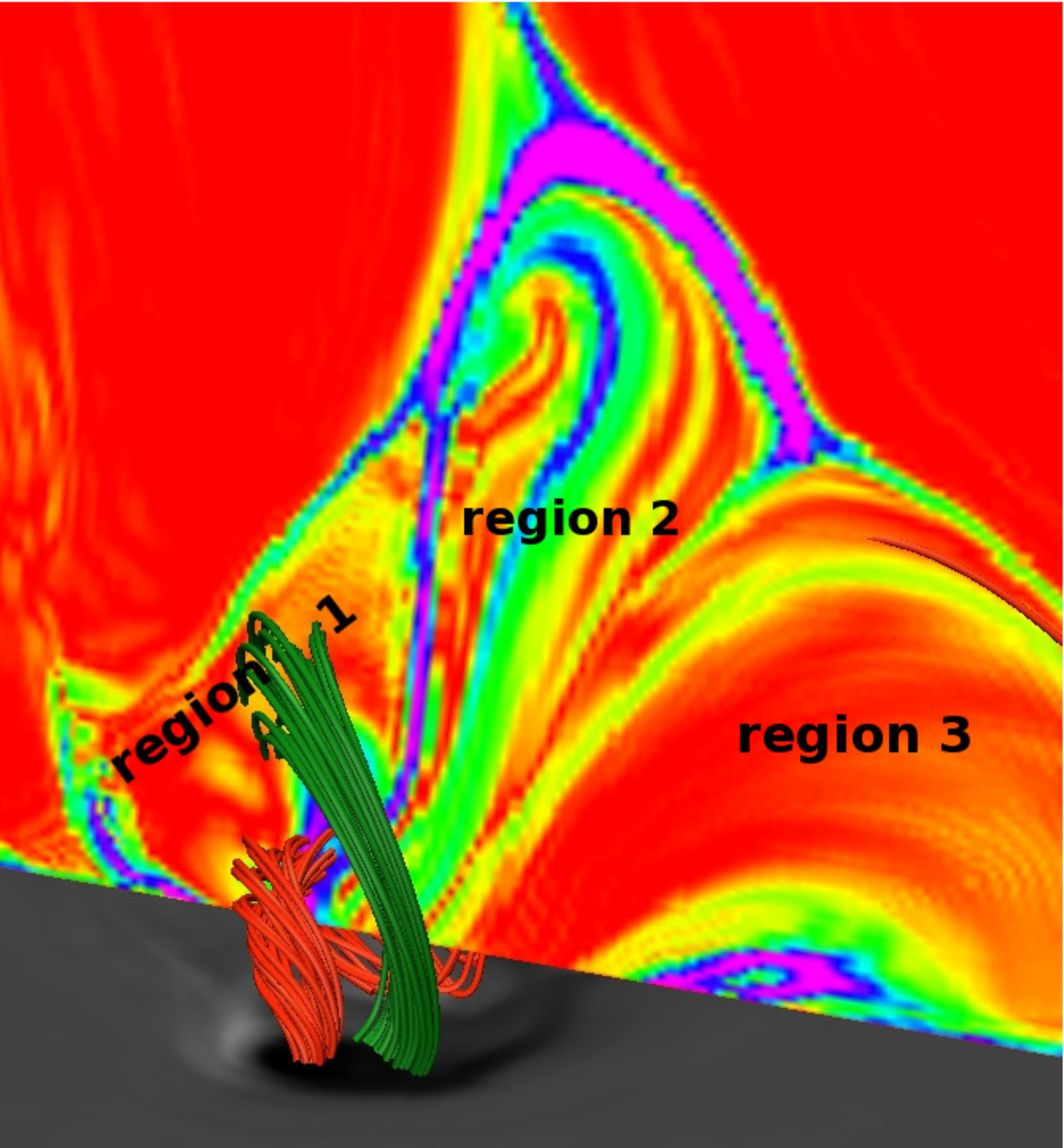}}
\vspace{4mm}
\centerline{\includegraphics[width=\sizefig]{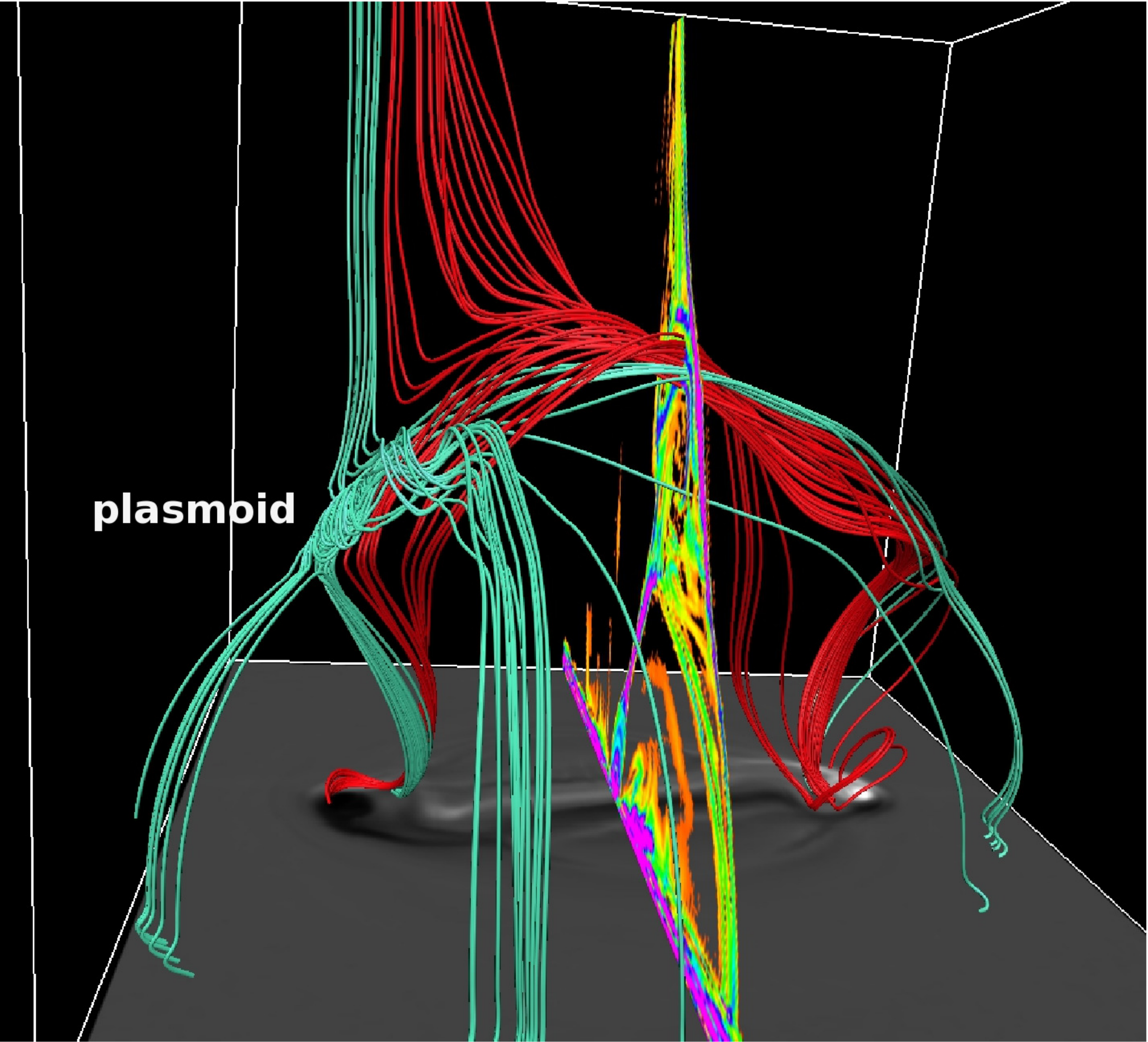}}
\vspace{4mm}
\centerline{\includegraphics[width=\sizefig]{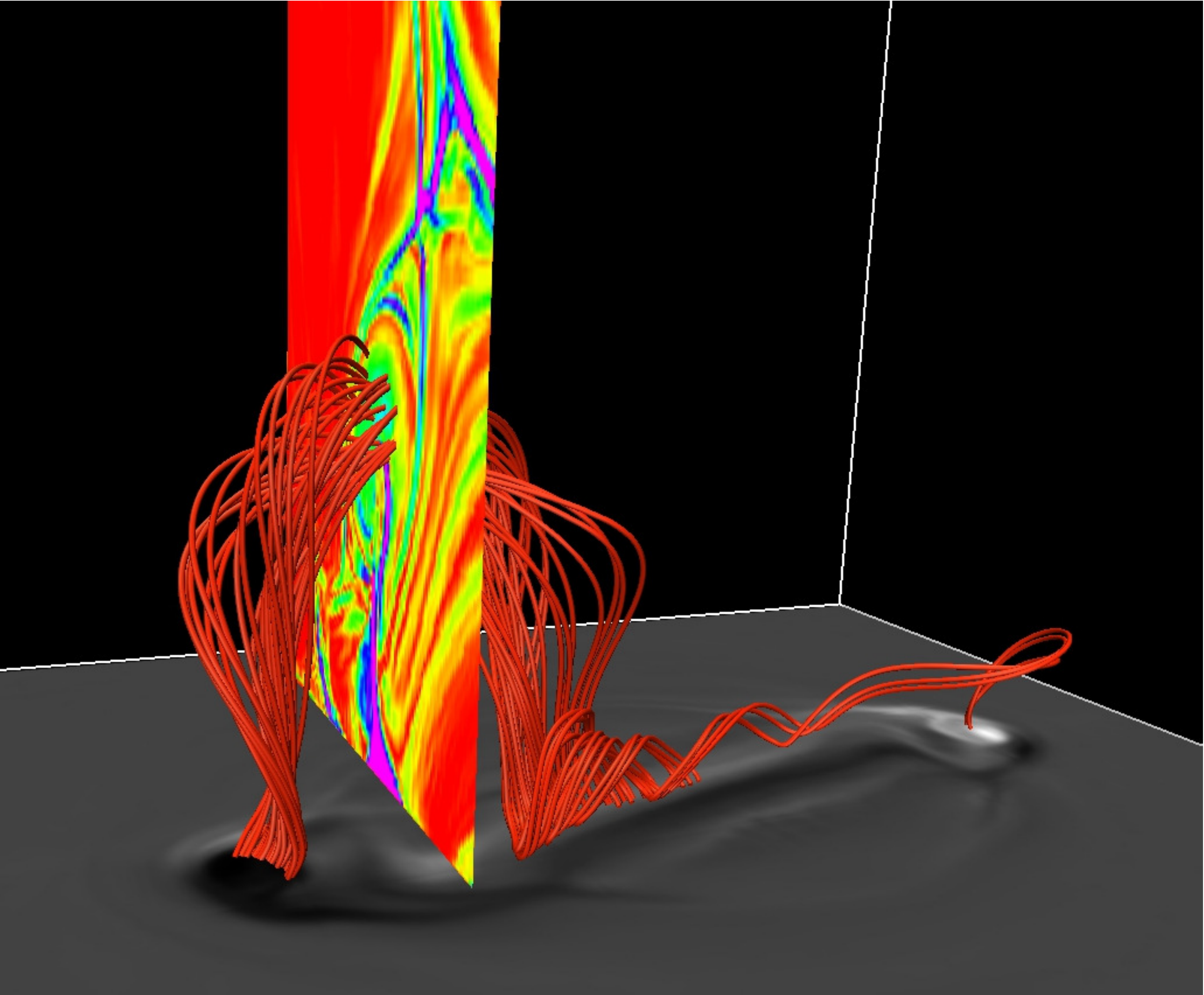}}
}
\hfill}
\caption[]{Top and center: Field line and electric current configuration at
  the start of the second eruption ($t=66$ min).  Top panel: map of $\joverB$
  at a vertical plane near the strong negative photospheric polarity ($y=6.3$
  Mm) including two flux ropes that erupt later on. Center: illustration of
  the field line topology associated with the prominent null point situated
  in the domain above the negative polarity. Bottom panel: rising twisted
  rope at $t=69.5$ min, illustrating the evolution of eruption number two in
  the experiment. The $\joverB$ vertical map shows an expanding and rotating
  bubble-like region which contains the erupting
  rope.} \label{fig:preeruption2}
\ifnum \value{twocol} > 0
\end{figure}
\else
\end{figure*}
\fi

\subsubsection{The second eruption}\label{sec:second_eruption}
As already indicated, in the region of the upward-pointing wedge
visible on the top panel of Fig.~\ref{fig:preeruption2}, the low-lying
arched and twisted flux rope is starting to move upward with
increasing speed, leading to the second eruption. The situation at an
intermediate stage of the process can be followed through the bottom
panel of the figure, drawn for $t=69.5$ min. There, we see a roundish,
bubble-shaped domain in the vertical plane pierced by a flux rope
(drawn in red) that has a set of footpoints rooted in the strong
negative flux concentration while the other end delineates the PIL as it gets
near the other polarity. The bubble is rising in that plane with
velocities in excess of $100$ km \persec\ .  The plasma attached to
the erupting rope has low temperatures, $1-4\,10^5$ K, and relatively
high densities compared with its surroundings, e.g., on the order of
$10^{11}$ \percc\ (in either case about one order of magnitude off the
values around the rope).  The rope is moderately twisted (say, between
$1$ and $2$ turns between the two footpoints), so the eruption
may follow the development of a kink instability in it. In support of
this hypothesis is the fact that the erupting rope appears to be
converting part of its twist into writhe, in some ways resembling the
evolution modeled by \citet{2005ApJ...630L..97T}; also the PIL
underneath the erupting rope gets increasingly hook-like. On the other
hand, the initial shape of the rope (as in the left panel) is more
like a semi-torus, so we cannot exclude that a torus instability
\citep{2006PhRvL..96y5002K} is also helping propel the
eruption. Finally, in the very early stages of this eruption there is
a current concentration right below the rope that corresponds to an
abrupt change of orientation of the field lines there. Hence,
a certain amount of reconnection below the rope may partially be
driving the eruption. The peak kinetic energy associated with this
eruption is quite similar to the peak kinetic energy associated with
the first one, namely of order $8\;10^{26}$ erg.

\subsubsection{The third eruption}\label{sec:third_eruption}

By time $t=74$ min, a third violent eruptive process takes place in the
emerged domain, this time in the positive surface-polarity end of the active
region, located in the $y<0$ semi-space. This eruption bears a number of
similarities with the second one, but is not fully equivalent to it since the
field configuration of the whole emerged domain has no left-right symmetry
with respect to the $y=0$ vertical plane. The initial configuration of the
field lines above the positive polarity is also of the twisted magnetic arch
type with a PIL underneath with hook-like appearance (see the top panels of
Figs.~\ref{fig:wedge} and \ref{fig:eruption_1_a_2}). Yet, the successive
reconnection processes that have already occurred have led to a global
connectivity pattern there which is quite different to the pattern above the
negative polarity. For instance, there is no prominent null point above the
positive polarity with fan-surface and spines extending for a long distance,
as we saw for the negative polarity in
Sec.~\ref{sec:null_above_negative_polarity}. Also, like before, by drawing a
map of $\joverB$ on a vertical plane one can see a three-fold arrangement of
connectivity regions, but, now, the unperturbed, twisted flux rope that winds
its way above the PIL close to the strong photospheric polarity is found in
the central region between the perturbed region and the hot loops and not on
one side as was the case for the second eruption.

The external appearance of this eruption is not very different to the
previous one: the unperturbed twisted flux rope around the strong
photospheric polarity gets ejected upward, whereby it appears to
convert part of its twist into writhe along the rise, as deduced both
from the general shape of the rising arch as well as through the
increasingly winding shape of the PIL at the photosphere. The erupting
field lines have a sufficient number of turns around the rope axis to
make it likely that the eruption is associated with the development of
a kink instability. The strong perturbation rises and impacts on the
ambient coronal field above the emerged domain. As for the two
previous eruptions, the ambient open field is strong enough to
withstand the impact: the perturbation is channeled along the inclined
coronal field lines and creates a jet-shaped perturbation that reaches
the top of the numerical domain. Concerning the kinetic energy
associated with this eruption, there is no clean separation in the
energy curve between eruption 2 and 3: during the latter, the kinetic
energy in the volume goes to a maximum of about $1.1\;10^{27}$ erg.

\section{Time evolution of global quantities}
\label{sec:time_evolution_global}

\subsection{Height distribution of the axial magnetic flux} 
\begin{figure}
\hbox to \hsize{\hfill
\includegraphics[width=9cm]{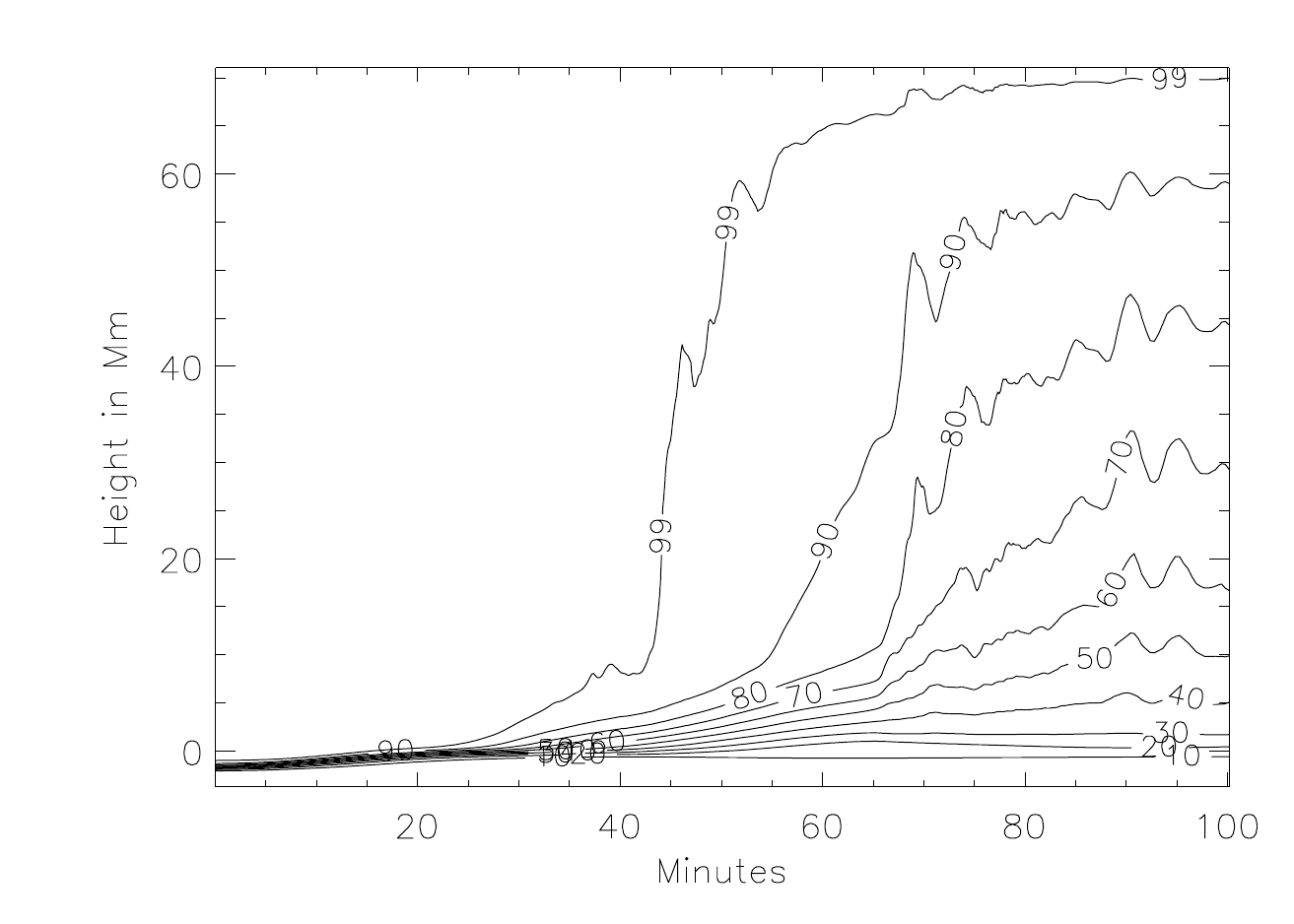}
\hfill}
\caption[]{\label{flux_height.fig} Redistribution of the $B_y$
  flux in time. Each curve corresponds to the height below
  which a given fraction of the total axial flux is contained (the fraction
  is annotated in each curve). }
\end{figure}

At time $t=0$ the ambient coronal field is contained in the $xz$
vertical planes, so it is orthogonal to the axis of the embedded
subphotospheric magnetic tube, which points in the $y$ direction. This
implies that any excess \hbox{$B_y$-\,flux} later found in the
atmosphere has to result from the emergence of the initial flux
tube. To investigate this process we calculate the flux across the
central plane $y=0$ between the bottom of the box and a given height:

\begin{equation}
\Phi_y(z,t) = \int_{z_{bot}^{}}^{z} dz' \int_{-x_{bound}^{}}^{x_{bound}^{}} dx' \, B_y(x',y=0,z',t)\;,
\end{equation}
with $x_{bound}$ the value of $|x|$ at the position of the $x$-boundary. 
To make a simple representation, we normalize
$\Phi_y(z,t)$ with the total flux across that plane (which is a
constant), and represent the inverse function, i.e., the height at
which a given fraction of the total flux is reached. 
Fig.~\ref{flux_height.fig}  shows the inverse function for percentages 
in $10\%$ steps: the curve labeled '90', for
instance, shows the height below which $90\%$ of the initial flux is
contained; the one below corresponds to $80\%$ and so on. 

We see how the flux tube initially ($t \lesssim 25$
  min) rises very 
slowly towards the photosphere. Thereafter, it starts its
expansion into the corona, first gradually and then at a faster
speed. Yet, until  $t\approx 65$ min only a small fraction of the
axial flux reaches high into the corona.  By that time the situation
changes dramatically and axial flux is expelled in a short burst into
the coronal domain; this corresponds to the first eruptive event
of Sec.~\ref{sec:advanced_phases}.  After that event,
roughly $50\%$ of the total axial flux is found above $10$ Mm over the
photosphere, which shows how effective the eruption is in pushing
axial flux far up into the coronal region in comparison to the
previous standard jet phase. In the later phases, the
spreading of flux slows down and the curves show an oscillatory behavior, most
likely caused by the non-equilibrium ensuing from the rapid
ejection. Notice that in this plane ($y=0$) the following eruptions do
not leave any major imprint in terms of redistribution of axial flux,
which is a further indication that they take place mostly in the
flanks of the active region.

\subsection{Connectivity between the submerged field and the
  atmosphere. The unsigned photospheric flux} 

The field lines of the initial submerged tube connect the two vertical
$y=$const planes at the boundary of the box with each other. As the tube
surges into the atmosphere, part of those initial field lines reconnect with
the ambient coronal field. We investigate the change in connectivity of the
initial tube by tracing at different times a large number of magnetic field
lines from a fixed circular area on one of the $y$-boundaries
(hereafter called {\it the initial circle}): this
circle covers the location of the initial submerged flux tube and a further
area around it so that at later times it still covers
most of the flux tube as it slowly diffuses out and
sinks. The connectivity of the field lines starting from that circle can be
divided into two groups: one (the {\it tube flux} for
  short) that connects the circle with a similar circle (but with 50\% larger
  radius) at the opposite $y-$boundary (the {\it target circle}), and then
the complementary set, consisting mostly of field lines
  that connect to the open field in the corona (here called {\it open
    flux}). Choosing a larger radius for the target
  circle helps prevent misidentifications of the connectivity class because
  of field line integration inaccuracies.
 
\begin{figure}[h]
{
\centerline{\hfill\includegraphics[width=0.5\textwidth]{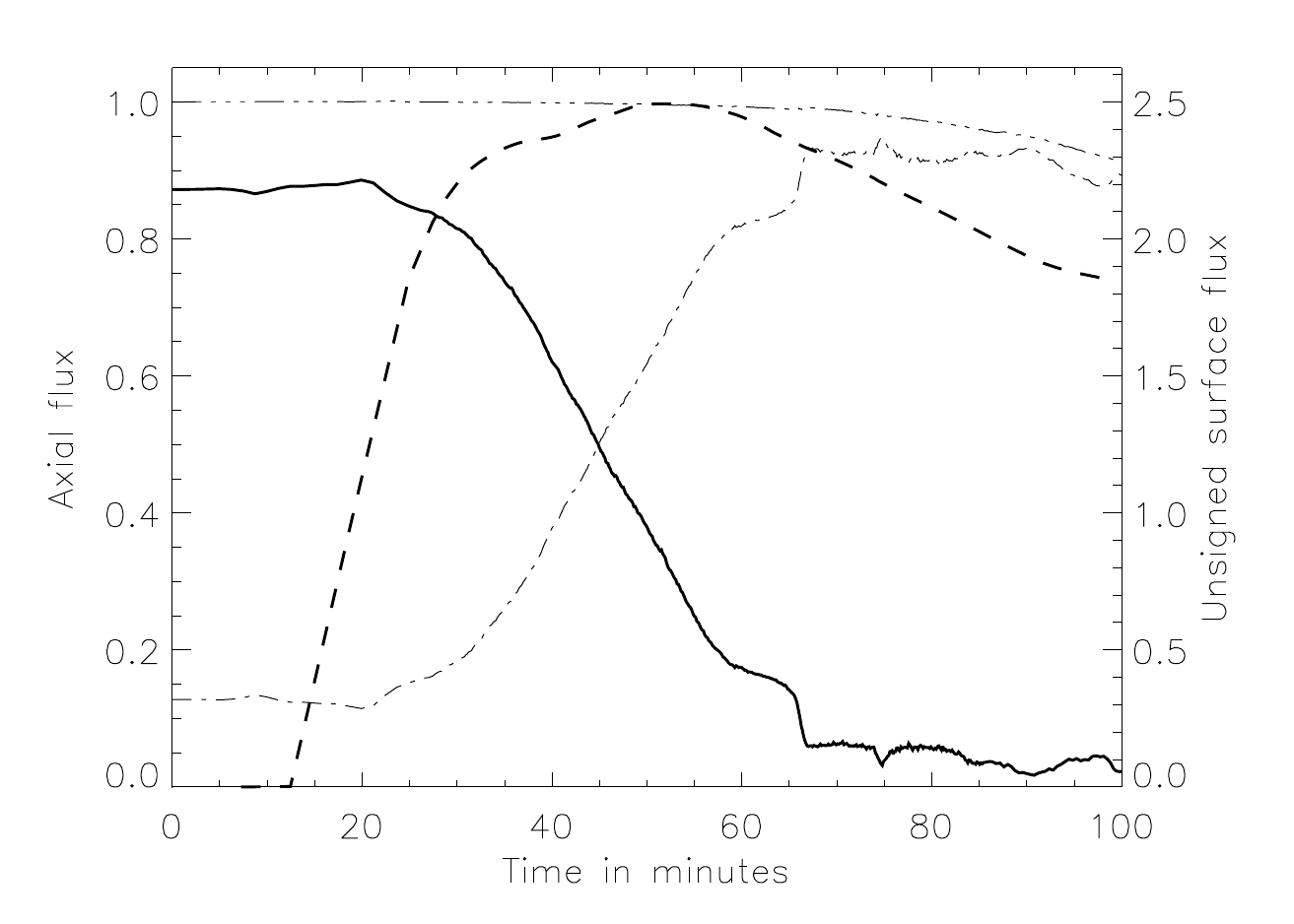} \, \hfill}}
\caption[]{\label{connectivity.fig} Change in time of the amount of flux
  connecting the two subphotospheric roots of the magnetic tube (left-hand
  ordinate axis).  The curves represent the flux connecting the roots
  (solid), the open 
  flux (i.e., the flux with other connectivities, dash-dotted) and their sum
  (dash-dot-dot), all normalized to the initial value of the sum. The
  thick dashed curve shows the unsigned vertical magnetic flux
  of either polarity through the photosphere, normalized to the initial axial
  flux in the tube (right-hand ordinate axis).  }
\end{figure}

\Fig{connectivity.fig} helps quantify the connectivity changes along
time. The full and dash-dotted lines give the fraction of the magnetic flux
in the initial circle that belongs to the {\it tube}- and {\it open}-flux
connectivity classes, respectively; the dash-dot-dot line gives their sum
which is not constant in time since a small amount of
  flux diffuses or is advected out of the initial circle). We first see that
at $t=0$, most ($90\%$) of the flux in the initial circle is part of the
tube.  The solid and dash-dotted curves are roughly horizontal until $t=20$
minutes, with the flux thereafter changing connectivity class in a monotonic
fashion until $t=58$ minutes. This is followed by a much slower decline of
the solid curve (with a mirrored pattern in the dash-dotted one)
until $t=63$ minutes, where a rapid and relatively small change occurs as a
result of the first eruption.  Toward the end of the experiment only a small
amount of flux is still connecting the two ends of the flux tube.  Seen from
the surface, we expect the strong bipolar flux concentrations at the
photosphere to lose their initial mutual connectivity and instead end up
connecting almost entirely to the ambient flux in the corona.

The thick dashed curve in Fig.~\ref{connectivity.fig} provides related
information: the curve shows the integrated unsigned magnetic flux through the
photosphere with the flux of the initial coronal field subtracted, the whole
being divided by two:
\begin{equation}
{1\over 2}\left(\int_{z=0} |B_z(x,y,t)| \,dx\,dy - 
\int_{z=0}^{}|B_z(x,y,t=0)| \,dx\,dy\right)\;.
\label{eq:flux}
\end{equation}

Given that most of the emerged flux at the photosphere is in the strong-field
concentrations, where $|B_z| >> |B_z(t=0)|$, that quantity approximates well
the total emerged flux of either polarity at the surface.  
The thick dashed-line of Fig.~\ref{connectivity.fig} shows Eq.~\ref{eq:flux}
in units of the axial flux of the initial tube.  The 
curve shows the arrival of flux at the surface ($t
  \approx 10$ min) and the rapid subsequent rise to a value well above the
initial axial flux due to the contribution of the
azimuthal components of the field in the rising
twisted tube. There is a good correspondence in time between the top region
of the dashed line and the phase of rapid change of connectivity shown by the
full and dash-dotted lines. Finally, from $t \approx 55$ onward, the total
unsigned surface flux decays slowly, and the eruptions and late episodes in
the corona have no counterpart in it: this is expected since the changes in
the coronal field are in no position to modify the flux in the photosphere.

As a final remark, note that, given the complication of the topological
changes for the field in the box, \Fig{connectivity.fig} cannot be used to
derive a global rate of reconnection between the emerging tube and the
coronal field, but, at best, a lower limit for it (see the related discussion
by \citealt{2008ApJ...675.1656P}).

\section{Discussion}
\label{sec:discussion}

\subsection{Comparison with jet observations: standard jets and blowout jets}

There are various remarkable qualitative and quantitative similarities
between our results and those of a number of recent observational papers. We
start by considering the values of the plasma temperature and density for the
jets mentioned in the introduction. The range of temperatures obtained by
\cite{2000ApJ...542.1100S} fit well with the temperatures obtained by us for
a generic height in the jet (Fig.~\ref{fig:max_vel_temp}), while the values
of \cite{2012A&A...545A..67M} are rather on the low side compared with
ours. During its hottest phase, the jet in the experiment also has low
densities, as explained in Sec.~\ref{sec:the_jet}. At that time we measure
densities between about $10^9$ \percc\ and a few times that value depending
on the position along the jet, which match well the values quoted by those
authors. In any case, note that
  Figs.~\ref{fig:max_vel_temp} and \ref{fig:time_evolution_in_current_sheet}
  contain peak values of $T$ and $v$, and that (see
  Fig.~\ref{fig:jet_horizontal_cuts}) the very fast jet strands may 
  be small in terms of the resolution element of the satellite instruments: a
  good comparison with observations requires a much deeper study than
  is possible within this paper.

In the introduction we also mentioned the statistical study of
Hinode X-ray polar jets by \cite{2007PASJ...59S.771S}.
The velocities they obtain, distributed between $70$ and about $500$ km
\persec\ and peaking at $160$ km \persec\ fit well with our simulations
(Figures~\ref{fig:max_vel_temp} and
\ref{fig:time_evolution_in_current_sheet}), in which we obtain values from
$100$ to $300$ km \persec\ ($400$ km \persec\ in the current sheet) along the
main phase of the jet. Concerning the observed longitudinal sizes (between
$10$ and $120$ Mm), we note that the jet in the numerical experiment easily
reaches the top boundary placed at a height of $70$ Mm, so the expectation is
that the jet motion will continue to substantially larger heights. The peak
height in the observation may be indirectly limited by the weakness of the
signal as the plasma expands with height. For the width of the jet,
\cite{2007PASJ...59S.771S} mention some $8$ Mm; for the comparison, we use
our Fig.~\ref{fig:jet_horizontal_cuts}, from which we deduce a transverse
size of about $10$ Mm.  The lifetime of the jets was between $5$ and $30$ min
in the study of \citeauthor{2007PASJ...59S.771S}, peaking at $10$
min. Referring to our Fig.~\ref{fig:max_vel_temp}, we see our jet to last
some $30$ min, i.e., toward the upper bound of the observations. However, it
is unclear for which fraction of that time the jet would have sufficient
emission in the X-Ray range to be detectable by the XRT instrument. Finally,
the jet is seen to have a drift velocity in the horizontal direction as a
consequence of the reconnection process (Sec.~\ref{sec:hot_loops}). From the
observations the drift velocity is found to be on the order of $0-35$ km
\persec. In our experiment we have some $8$ km
  \persec\ for the drift velocity at maximum (end of
  Sec.~\ref{sec:hot_loops}). All in all, the comparison between theory and
the statistics of the observations is satisfactory: the present experiment
provides values for the different magnitudes (well) within the ranges
observed for actual X-ray jets in the Sun.

Of particular interest is the comparison with the recent observational results
concerning the so-called {\it blowout jets}. Using Hinode/XRT and Stereo/EUVI
observations, \citet{2010ApJ...720..757M} deduced a dichotomy of coronal jet
evolutionary patterns. Those authors detected that an important fraction of
the jet events, around one third of them, presents an eruptive pattern
accompanying the jet ejection itself. For this class of jets they find
anomalous X-ray brightenings below the jet additionally to the major bright
points of standard jets, followed by the eruption of material with
temperatures one or two orders of magnitude below coronal values. 
Further
observations of jets with associated energy release bursts are those of 
\citet{2011A&A...526A..19M} and \citet{2011ApJ...735L..18L}. In the
first of those, in particular, the author presented observations of a jet
with four successive short-duration energy release bursts over a time period of $20$ minutes.

The numerical experiment we are presenting in this paper yields in a natural
fashion a collection of eruptive phenomena directly associated with the
magnetic configuration around a jet following flux emergence which may be the
counterpart to these observational facts. We have found ejections of,
basically, two different kinds that we tentatively ascribe to (a) a
tether-cutting reconnection pattern occurring in the coronal remnant of an
emerged magnetic region and taking place in the complicated topology left
over by the reconnection that led to the jet in the first place; and (b) a
secondary twisted flux rope ejection pattern that may be caused
by the development of a kink instability (or perhaps of
  a torus instability). For the identification with observed features it may
be of interest to note that the eruptions in the experiment first go through
a direct impulsive phase with high velocities oriented in the outward
direction, later yielding a more chaotic velocity pattern when the eruption
impinges on the surrounding plasma, especially on the ambient coronal field,
which in our experiment is strong enough to divert the outgoing
flow. Analyzing the second eruption in our experiment
(Sec.~\ref{sec:second_eruption}), we have seen that the ejected flux rope
contains cool ($1-4\; 10^5$ K) and dense (almost $10^{11}$ cm$^{-3}$)
material during the initial impulsive phase, values which differ by about one
order of magnitude from those of the surroundings. A detailed analysis of
further aspects of this multi-faceted phenomenon is necessary to ascertain 
its possible identification with the results of \citet{2010ApJ...720..757M}.

\subsection{The eruptions and their relation to previous experiments}

In the numerical simulation literature there are different reports of
eruptions resulting directly or indirectly from an episode of flux emergence
into the atmosphere \citep[e.g.,][]{2003ApJ...589L.105F,2004ApJ...609.1123F,
  2004ApJ...610..588M, 2008A&A...492L..35A, 2008ApJ...674L.113A,
  2012A&A...537A..62A, 2009A&A...508..445M, 2009ApJ...704..485T}, 
or, also, from processes related
with the development of a kink- or torus instability in a magnetic
configuration \citep{2003A&A...406.1043T, 2005ApJ...630L..97T,
  2009ApJ...691...61P, 2010ApJ...714.1762P, 2010ApJ...708..314A}.
\cite{2004ApJ...610..588M} followed the emergence of a flux tube into a
nonmagnetic coronal domain. They explain how after an initial phase of
emergence from the interior, Lorentz-force driven motions 
lead to the conversion of the emerged magnetic domain into a sheared arcade;
its upper part  starts to separate from the lower part,
which is loaded with mass and must stay behind.  The process results in local
stretching of the field lines in the vertical direction just above the
photosphere 
and formation of a vertical current sheet above the polarity inversion
line. As the current builds up in strength, a fast reconnection process is
initiated in the arcade.  This rapidly creates a new flux rope above the
current sheet that moves upwards with a speed much higher than the rise
speeds in the previous phases and constitutes an eruption. \cite
{2012A&A...537A..62A} expanded this work by conducting experiments
with different ambient coronal field configurations.
They found the same eruptive behavior of the emerging flux tube independently
of the coronal field; they also showed that the dynamics following the
eruption depends on both the twist of the initial tube and the orientation
of the imposed coronal field. Depending on the parameter combination, the
erupting flux tube will either escape up through the atmosphere or be
confined by the overlying coronal magnetic flux.

The eruptions discussed in those investigations are comparable to our first
eruption. However, given the previous evolution of the system in our case,
with four connectivity patterns resulting from the emergence of the initial
submerged rope into a slanted coronal field
(Fig.~\ref{fig:3Dview}), we had 
to ascertain the complicated changes of connectivity of different
regions caused by the eruption (Sec.~\ref{sec:first_eruption}) and, also,
determine regions in the emerged volume which remain unaffected by it
and, in fact, turn out to be ejected themselves later on (as in eruptions
2 and 3). We conclude that the standard sheared arcade model requires to be
modified to deal with the more complex field line situation found in our
experiment. The additional complexity also allows information from the
reconnection process to spread over a larger volume and therefore indirectly
stress neighboring flux regions that would not be reached in the classical
picture.

In a different approach to simulate jet formation
\cite{2009ApJ...691...61P,2010ApJ...714.1762P} conducted a series of
experiments in which a vertical dipole is embedded in a background monopolar
field. This gives a magnetic topology in which a 3D null point hovers above
the parasitic polarity region with a near vertical spine axis that in one
direction reaches the model photosphere close to the center of the parasitic
polarity and in the other follows the open field upwards.  Stress is imposed
in the form of a systematic rotation of the parasitic polarity region. The
twisting of the spine axis eventually destroys the symmetries in the system
and drives a kink instability. The evolution of the null point is, in a way,
comparable to what we found for the null point near the site of the second
eruption -- disrupting the fan plane by building up a current sheet that
splits the external part of the spines' axis as also
seen in dedicated null point investigations (see
\citealt{2011A&A...534A...2G} and references therein; see also
\citealt{2009ApJ...704..485T, 2009ApJ...700..559M}
for other experiments with driven fan-spine reconnection events.)
However, from Sec.~\ref{sec:later_eruptions}, it is
  clear that it is not a kink instability associated with the rotation of the
  spine axis
that is responsible for the second and third eruptions. Instead we think
that in our case it is a twisted rope with the shape of a highly inclined
$\Omega$ loop at the end points of the PIL that becomes unstable and erupts,
thereby converting twist into writhe. So we assume that the driver for the
eruption is more of the kink- or torus-instability kind of a semitoroidal
twisted rope, perhaps a combination of the two.

\subsection{Flux emergence, H$\alpha$ surges and CaII jets}

In this experiment we have found that a substantial amount of cool and dense
material accumulates around the main dome of emerged material right before
and during the main phase of jet ejection; the accumulated material has
densities at a given height typically one to two orders of magnitude above
those at that height in the unperturbed initial corona
(Sec.~\ref{sec:cold_region}, Fig.~\ref{fig:density_wall_hor_cut}). The
maximum accumulation occurs in the domain below the jet, where it can reach a
vertical extent of several Mm.
Through extended Lagrange tracing we have determined that a
relevant part of the plasma in that cool and dense wall originally was in
the emerged magnetic dome and has been bodily transferred to the wall
accompanying field lines that change connectivity at some height, but the
plasma element itself need not have gone through the reconnection site. Other
plasma elements in the wall were originally located at comparatively low
levels in the ambient corona when the flux tube arrived, from where they were
transported to the wall, again in many cases accompanying field
lines reconnecting somewhere along their length. The velocities measured in
the dense domain are not large, typically less than $50$ km \persec, and the
temperatures range between a few times $10^4$ K and a few times
$10^5$ K, i.e., mostly transition-region values.  These cool features do not
have any obvious jet appearance; in particular, from the morphology of the
distributions of velocity, temperature or density, one cannot say that they
contain, or consist of, collimated motions. It is also unclear
what the fate of this cool region would be if heat conduction had been
included in the experiment.

Now, a substantial amount of observational and theoretical literature has
been devoted to the quasi-simultaneous appearance of hot X-ray/EUV jets and
cold H$\alpha$ surges or CaII-H jets \citep[see,
  e.g.,][]{1995SoPh..156..245S, 1996ApJ...464.1016C, 1999ApJ...513L..75C,
  1999SoPh..190..167A, 2007A&A...469..331J}. A theoretical interpretation has
been based on the possible simultaneous presence of cold and hot jets in
two-dimensional numerical experiments of flux emergence into magnetized
coronae (\citealt{1996PASJ...48..353Y, nishizuka_etal_2008}; see also
\citealt{2013arXiv1301.7325T}). \citet{1996PASJ...48..353Y}
  state that the cool plasma is chromospheric in origin and is, using the
  words in that paper and in \citet{nishizuka_etal_2008}, {\it ejected or
  accelerated by the sling-shot effect due to reconnection, which produces a
  whip-like motion}.  The relation of this description for the 2D experiments
  with the 3D processes analyzed in our Sec.~\ref{sec:cold_region} needs
  further exploration.  \citet{nishizuka_etal_2008} also suggest that the
cool jet may result from the asymmetry of the plasma properties on either
side of the main current sheet separating the ambient corona (hot and thin
plasma) from the emerged domain (cool and dense): this would lead to the
formation of both a fast hot jet and a cool, slower
jet. In any case, in view of the available theoretical
  and observational results, we think that a clear identification of the cool
  structures in 2D or 3D with observed cool jets does certainly require a
  more in-depth comparison between simulations and observations.

\subsection{Summary}

We have studied the fast and hot jet and the violent eruptions that follow
the emergence of organized magnetic flux from the solar interior into a corona
with a uniform, inclined magnetic field. Here is
   a summary of some major results:

\begin{itemize}
\item The quantitative and qualitative aspects of the 3D
  structure and evolution of the {\it standard jet} that is launched
  following the collision of emerging and ambient field systems have been
  explained (Sec.~\ref{sec:jet_reconnection})
\item In the standard jet phase, the reconnection process that takes place at
  the interface between the colliding flux systems has been seen to be
  topologically complex, including both magnetic null points and
  plasmoids. This results in the formation of complicated connectivity
  patterns, and spatial and temporal variation of physical variables
  (Sec.~\ref{sec:current_sheet_reconnection}).
\item An indirect product of the reconnection process in the standard jet phase 
  is the formation of a dense and cool 3D plasma structure surrounding the emerged 
  plasma region (Sec.~\ref{sec:cold_region}).
\item The different physical parameters of the standard jet resulting
  from our experiment are well within the observational ranges obtained in the
  statistical study of \citet{2007PASJ...59S.771S}.
\item After the standard jet phase, a number of violent eruptions of the
  magnetic field structure take place in different volumes of the emerged
  plasma dome. They show a variety of origins and patterns of evolution.
  This covers the traditional sheared arcade instability as described by
  \cite{2004ApJ...610..588M}, but also other types that are probably 
  related to the development of a kink or torus instability.
\item The global scheme resulting from this
  experiment (flux emergence, standard jet launching and violent eruptions of
  different physical origins) may provide an explanation for the phenomenon
  of blowout jets described by \citet{2010ApJ...720..757M}. 
\end{itemize}

\acknowledgements{ Financial support by the Spanish Ministry of Research and
  Innovation through projects AYA2007-66502, AYA2011-24808 and CSD2007-00050
  and by the European Commission through the SOLAIRE Network
  (MTRN-CT-2006-035484) is gratefully acknowledged.  We are grateful to ISSI
  (Berne, Switzerland) for hosting two series of workshops (on magnetic flux
  emergence and on coronal jets) where parts of this work were presented.
  The results in this paper have been achieved through generous computing
  time grants by the European Consortia PRACE (JUGENE and JUROPA, J\"ulich)
  and DEISA (HLRS, Stuttgart), as well as by BSC (MareNostrum, National
  Supercomputing Center, Barcelona), GCS/NIC (Danish Center for Scientific
  Computing, Copenhagen), and IAC (LaPalma supercomputer, Canary Islands):
  their permission to use the resources and their support are greatly
  appreciated. 3D visualisation has been carried out using the VAPOR package
  (UCAR, www.vapor.ucar.edu). The useful comments by an anonymous referee are
  gratefully acknowledged.  We are also grateful for many conversations with
  M.~Madjarska on the observations and the physics of coronal jets as well as
  for comments on the  manuscript by I.~Ugarte Urra.}

\bibliography{references}

\end{document}